\documentclass[fleqn,usenatbib]{mnras}

\usepackage{newtxtext,newtxmath}

\usepackage[T1]{fontenc}
\usepackage{ae,aecompl}

\usepackage{graphicx}	
\usepackage{amsmath}	
\usepackage{amssymb}	
\usepackage{soulutf8} 
\usepackage[utf8]{inputenc} 
\usepackage{enumerate} 
\usepackage{color} 

\usepackage{multicol}
\usepackage{hyperref} 
\usepackage{epstopdf}

\definecolor{wildwatermelon}{rgb}{0.99, 0.42, 0.52}
\definecolor{trolleygrey}{rgb}{0.5, 0.5, 0.5}
\definecolor{cobalt}{rgb}{0.0, 0.28, 0.67}
\definecolor{bleudefrance}{rgb}{0.19, 0.55, 0.91}
\definecolor{cerulean}{rgb}{0.0, 0.48, 0.65}
\definecolor{mspgreen}{rgb}{0.13, 0.55, 0.13}
\definecolor{orange}{rgb}{0.6,0.4,0.0}
\definecolor{denim}{rgb}{0.73, 0.2, 0.52}

\newcommand\heii{\ion{He}{ii}~4686~\r{A} }
\newcommand\hheii{\ion{He}{ii}~4686~\r{A}}
\newcommand\hei{\ion{He}{i}~4921~\r{A}}

\newcommand\vp{V348~Pav\ }
\newcommand\vvp{V348~Pav}
\newcommand\cyc{{\sc cyclops}\ }

\title[Observations and modelling of the polar V348~Pav]{Optical observations and \bfseries{\scshape{cyclops}} post-shock region modelling of the polar \vp}

\author[A. S. Oliveira et al.]{Alexandre S. Oliveira,$^{1}$\thanks{E-mail: alexandre@univap.br}
Claudia V. Rodrigues$^{2}$,
Matheus S. Palhares$^{1}$,\newauthor
Marcos P. Diaz$^{3}$,
Diogo Belloni$^{2,4}$,
Karleyne M. G. Silva$^{5}$
\\
$^{1}$ IP\&D, Universidade do Vale do Paraíba, Av. Shishima Hifume, 2911, 12244-000, São José dos Campos/SP, Brazil\\
$^{2}$ Instituto Nacional de Pesquisas Espaciais, Av. dos Astronautas, 1758, São José dos Campos/SP, 12227-010, Brazil\\
$^{3}$ IAG, Universidade de São Paulo, Rua do Matão, 1226, 05508-090, São Paulo/SP, Brazil\\
$^{4}$ Instituto de F{\'i}sica y Astronom{\'i}a, Universidad de Valpara{\'i}so, Av. Gran Breta{\~n}a 1111 Valpara{\'i}so, Chile\\
$^{5}$ European Southern Observatory, Alonso de Córdova 3107, Vitacura, Santiago, Chile
}

\date{Accepted XXX. Received YYY; in original form ZZZ}

\pubyear{2019}

\begin{document}
\label{firstpage}
\pagerange{\pageref{firstpage}--\pageref{lastpage}}
\maketitle

\begin{abstract}

Post-shock regions (PSR) of polar cataclysmic variables produce most of their luminosity and give rise to high circular polarization in optical wavelengths and strong variability on the white dwarf rotation period, which are distinctive features of these systems. To investigate the polar candidate \vvp, we obtained a comprehensive observational set including photometric, polarimetric, and spectroscopic data, which was used to constrain the post-shock properties of the system. The object presents high circular polarization ($\sim30$ per cent) and high \heii to H~$\beta$ line ratio, confirming it is a polar. From both radial velocities and light curves, we determined an orbital period of 79.98~min, close to the orbital period minimum of CVs. The H$~\beta$ radial velocity curve has a semi-amplitude of $141.4\pm1.5$ km s$^{-1}$. Doppler tomography showed that most of the spectral line emission in this system is originated in the region of the companion star facing the WD, possibly irradiated by the emission related to the PSR. We modelled the PSR using the \cyc code. The PSR density and temperature profiles, obtained by a proper solution of the hydro-thermodynamic equations, were used in a 3D radiative transfer solution that takes into account the system geometry. We could reproduce the \vp \textit{B, V, R}, and \textit{I} photometric and polarimetric data using a model with a WD magnetic field of $\sim28$ MG, a WD mass of $\sim0.85$~M$_{\sun}$ and a low ($\sim 25\degr$) orbital inclination. These values for the WD mass and orbital inclination are consistent with the measured radial velocities. 

\end{abstract}




\section{Introduction}

Cataclysmic Variables, or CVs, are binary star systems composed of a white dwarf (WD) that accretes material from a low-mass 
companion. The mass transfer occurs via Roche lobe overflow and usually establishes an accretion disc around the WD. However, when the WD has a magnetic field larger than around $10^6$~G, the accreted gas flows along the magnetic field lines and an accretion column leads the material directly to the WD surface. These are the magnetic Cataclysmic Variables (mCVs), which are classified either as polars (also known as AM Her stars), whose magnetic fields are stronger than $ \gtrapprox 10^7$~G and prevent the formation of the accretion disc, or as intermediate polars (IPs) with $B \lessapprox 10^7$~G, where an internally truncated disc may exist. In polars, the post-shock region in the accretion column emits X-ray radiation by thermal bremsstrahlung, while the optical and infrared radiation is dominated by cyclotron emission. Objects of this class are the strongest sources of 
circularly polarized light, attaining a polarized fraction of up to 50 per cent in optical bands. A review on polars may be found in \citet{cropper1990}.

\citet{2017AJ....153..144O} conducted a project to identify specifically new mCVs among targets selected, by variability criteria, from catalogues of photometric transient surveys, like the Catalina Real-Time Transient Survey -- CRTS \citep{2009ApJ...696..870D}. The main purpose of that project was to improve the statistics of the classes of mCVs, increasing the number of members. Larger samples of polars and IPs are important, for instance, to address questions about the evolution of mCVs, in particular the evolutionary link between IPs and polars.
The project consisted of spectroscopic snapshots of the selected targets to search for spectral signatures of magnetic accretion, such as the intensity of the \heii relative to the H~$\beta$ emission line. The mCV candidates found in this spectroscopic survey are then subject to follow-up time-resolved observations, to confirm their classification and to characterise the detailed nature of the systems.

\vp (= V1956-6034 = SSS1956-60) is one of the mCV candidates selected from the sample in \citet{2017AJ....153..144O} for a detailed analysis. This object was first detected by its variability in the SRC-J plates and suggested to be an AM Her system by \citet{1994AJ....107.2172D} for the presence of the \heii (weak) emission line in its spectrum. Its low intensity (\heii/H~$\beta$ $\sim$ 0.25) at that occasion, though, cast doubt on the polar interpretation for the system. It is classified as a nova-like CV in \citet{2001PASP..113..764D}.
\vp was identified as a transient source by CRTS in 2011 May 26, when it received the ID SSS110526:195648-603430. In this paper, we present photometry, polarimetry and spectroscopy of \vvp. These data confirm that this object is a polar and allow us to provide a description of the system properties. In the next section we present our data. Section \ref{sec:photresults} shows the analysis of the photometric and polarimetric data. In Section \ref{sec:spec_results} we describe the spectroscopic data and the Doppler Tomogram of \vvp. The modelling of the cyclotron emission is presented in Section \ref{sec:cyclops}. We present the conclusions in Section \ref{sec:concl}.

\section{Observations and data reduction}
\label{sec:observations}

Photometric and polarimetric observations of \vp were performed at the Observat\'orio Pico dos Dias (OPD--LNA/MCTIC), located in south-east Brazil, using the 1.6-m Perkin--Elmer telescope, while spectroscopic data were obtained, simultaneously with the polarimetric observations in 2014, at the Southern Astrophysical Research (SOAR) Telescope, on Cerro Pachón, Chile. The journal of observations is presented in Table~\ref{obslog}. The data acquisition and reduction are described in the following sections.

\begin{table*}
\caption{Log of Observations.}
\label{obslog}
\begin{tabular}{llccccc}
\hline
Date & Telescope & Instrument & Filter  & Exp. time (s)  & Number of exps. & Average mag. \\
\hline
2014 Apr 28                  & OPD  1.6-m    &  Ikon CCD               & $V$     & 30  & 241  &   15.9  \\        
2014 June 21                  & OPD  1.6-m    &  Polarimeter + Ikon CCD & $B$    & 40         & 34   &  16.1 \\
2014 June 21                  & OPD  1.6-m    &  Polarimeter + Ikon CCD & $V$   &  30        &  144  & 16.0  \\
2014 June 22                  & SOAR 4.1-m    &  Goodman Spectrograph   & ...   & 300       & 92   &  ... \\
2014 June 22                  & OPD  1.6-m    &  Polarimeter + Ikon CCD & $B$     & 60   & 251   & 16.1 \\
2014 June 23                  & SOAR 4.1-m    &  Goodman Spectrograph   & ...   & 300       & 70  &  ... \\
2014 June 23                  & OPD  1.6-m    &  Polarimeter + Ikon CCD & $I$     & 40 / 30   & 176 /251 &  15.3 \\
2014 June 24                  & OPD  1.6-m    &  Polarimeter + Ikon CCD & $R$     & 30   & 505  & 15.3  \\
2014 July 20                  & OPD  1.6-m    &  Polarimeter + Ikon CCD & $B$    & 30   & 250  &  16.3  \\
2014 July 20                  & OPD  1.6-m    &  Polarimeter + Ikon CCD & $V$   &  30  &  256  &  16.2 \\
\hline
\end{tabular}
\end{table*}

\subsection{Photometry and polarimetry}
\label{sec:polarimetry}

In 2014 Apr 28, photometric measurements of V348~Pav were made in $V$ band with a thin, back-illuminated $2048\times2048$ E2V CCD (CCD Andor iKon-L936-BR-DD SN: 13739) with 13.5 \micron~pixel$^{-1}$.
In the remaining OPD runs, polarimetric data were obtained in $B$, $V$, $R$ and $I$ bands with the same iKon CCD, coupled to a polarimetric module, as described in \citet{1996ASPC...97..118M}. Accurate timings were provided by a GPS receiver. The data reduction (bias and flatfield corrections) followed the standard procedures using {\sc iraf}\footnote{{\sc iraf} is distributed by the National Optical Astronomy Observatories, which are operated by the Association of Universities for Research in Astronomy, Inc., under cooperative agreement with the National Science Foundation.}. Differential aperture photometry was performed with the {\sc daophot~ii} package for the photometric data. The polarization was calculated according to \citet{1984PASP...96..383M} and \citet{1998A&A...335..979R} using the package {\sc pccdpack} \citep{pereyra} and a set of {\sc iraf} routines developed by our group\footnote{http://www.das.inpe.br/$\sim$claudia.rodrigues/polarimetria/reducao\_pol.html}. Each set of eight images was used to produce one measurement of linear and circular polarizations. We grouped the images in the following way: 1--8, 2--9, 3--10, and so on. Hence the polarization points are not independent measurements. The polarization position angle correction to the equatorial reference system was performed using the standard polarized stars BD~144922, BD~125133, HD~110984, HD~111579, HD~155197 and HD~316232 \citep{1990AJ.....99.1243T,2007ASPC..364..503F}. The polarization of the non-polarized standards (WD~1620$-$391 and WD~2007$-$303 from \citealt{2007ASPC..364..503F}) is consistent with zero, therefore no instrumental polarization correction was applied. We could not determine the correct sign of the circular polarization -- if positive or negative -- but we could detect changes in this sign.
The ordinary and extraordinary counts of the polarimetric data were summed to obtain the total counts of each object. Hence the polarimetric data allowed us to perform differential photometry and to obtain the light curves of the object. 

All OPD light curves were calibrated in magnitude, using as reference the USNO-A2.0 0225-30444660 star \citep{1998AAS...19312003M}. Its SDSS magnitudes ($g'=14.054$ mag, $r'=13.483$ mag and $i'=13.424$ mag, \citealt{2015AAS...22533616H}) were transformed to $ugriz$ SDSS system\footnote{http://classic.sdss.org/dr7/algorithms/jeg\_photometric\_eq\_dr1.html\#usno2SDSS} and then converted to $BVRI$ magnitudes using Lupton's equations\footnote{http://classic.sdss.org/dr7/algorithms/sdssUBVRITransform.html\#Lupton2005}, resulting in $B=14.49$~mag, $V=13.73$~mag, $R=13.08$~mag and $I=13.02$~mag. The differential and absolute calibrations were also performed using other field stars, and all results were consistent. The absolute values of the magnitudes have an error of about 0.2~mag, but the differential errors are about 10 to 15 times smaller than that.

\subsection{Spectroscopy}
\label{sec:spectroscopy}

Spectroscopic time-series were performed  in remote observing mode, simultaneously with the polarimetric observations. We used the SOAR Telescope with the Goodman High Throughput Spectrograph \citep{2004SPIE.5492..331C} and a Fairchild $4096 \times 4096$ CCD with 15 \micron~pixel$^{-1}$. The CCD was configured with 2.06~e$^{-}$~ADU$^{-1}$ gain. The spectrograph was set to operate with the 1200 l mm$^{-1}$ grating and the 0.84-arcsec slit, resulting in 1.6 \AA~ FWHM spectral resolution in the 4320 -- 5620 \AA~ spectral range. Quartz lamps calibration flats, and bias images, were taken to correct the CCD signature, while CuAr lamp exposures were used for wavelength calibration, which resulted in typical 0.8 \AA~ or about 45 km s$^{-1}$ calibration RMS residuals. The [\ion{O}{i}] 5577 \AA~ telluric spectral line was used to assess the wavelength calibration accuracy. The slit was aligned to the parallactic angle to minimize light losses due to the differential atmospheric refraction. Observations of the LTT6248 spectrophotometric standard star \citep{1992PASP..104..533H} were used for flux calibration. The data reduction, spectra extraction and calibrations were performed with {\sc iraf} standard routines.

\section{Photometric and polarimetric results}
\label{sec:photresults}

\begin{figure}
\includegraphics[width=\columnwidth]{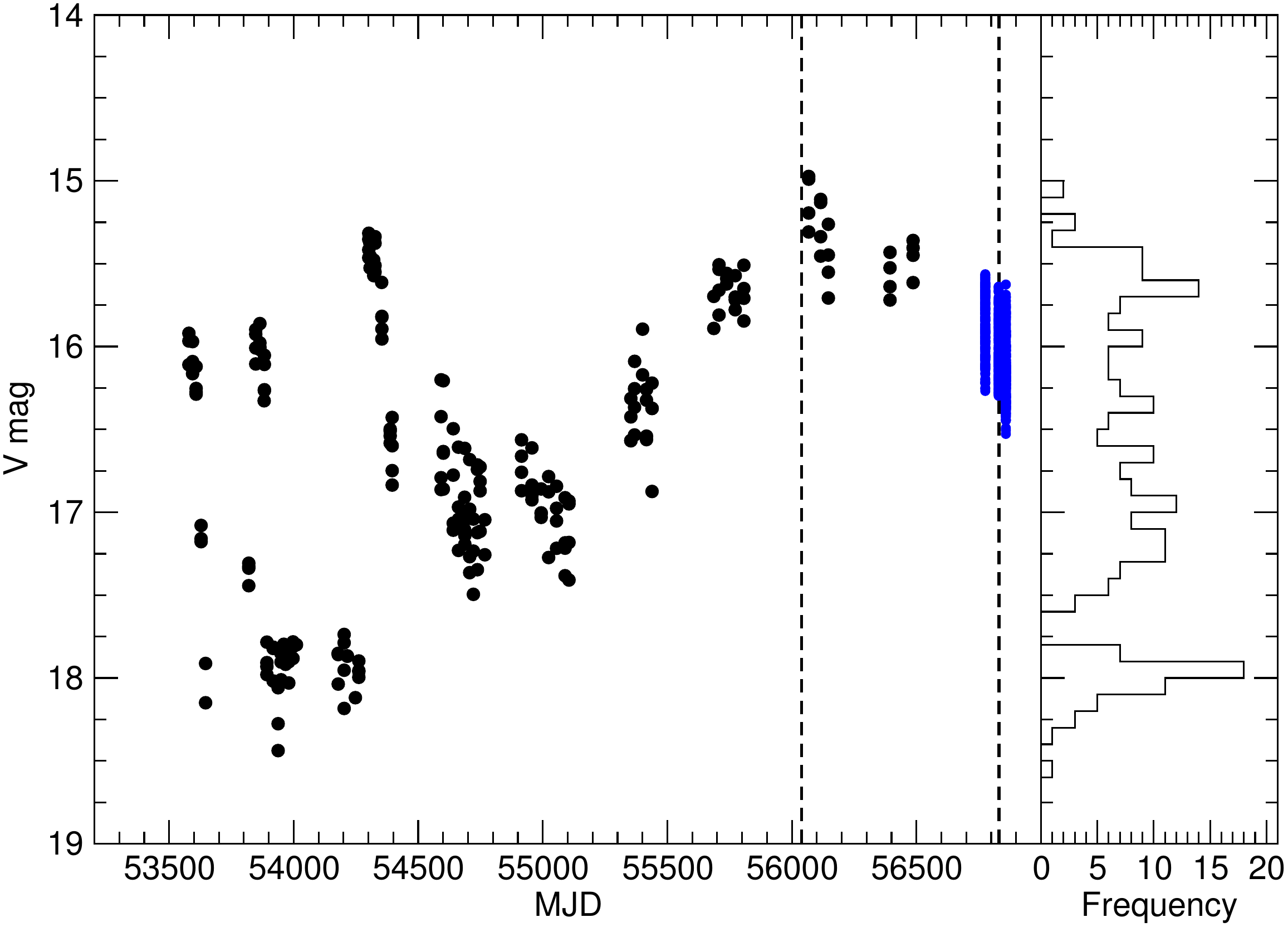}
\caption{Left panel: $V$-band CRTS light curve of \vp (black dots) and OPD $V$ filter measurements (blues dots). The vertical dashed lines indicate the date of our spectroscopic observations obtained in 2012 \citep[see][]{2017AJ....153..144O} and in 2014.  Right panel: Histogram of the CRTS magnitudes of \vvp. }
\label{fig:photoCRTS}
\end{figure}

In this section we describe the analysis of the $B$, $V$, $R$, and $I$ light and polarization curves obtained at OPD (see previous section), as well as the $V$ data available from CRTS. The CRTS dataset consists of 229 data points from the SSS survey, which span 8 years of observation (from July 2005 to July 2013) with a typical cadence of 4 consecutive exposures taken at intervals of 2 to 30 days (Fig.~\ref{fig:photoCRTS}). It presents variability between magnitudes 15 and 18.5 in time-scales of months, as well as up to 1 mag variations in time-scales of few days. 

\begin{figure*}
\begin{multicols}{2}
 \includegraphics[width=\linewidth]{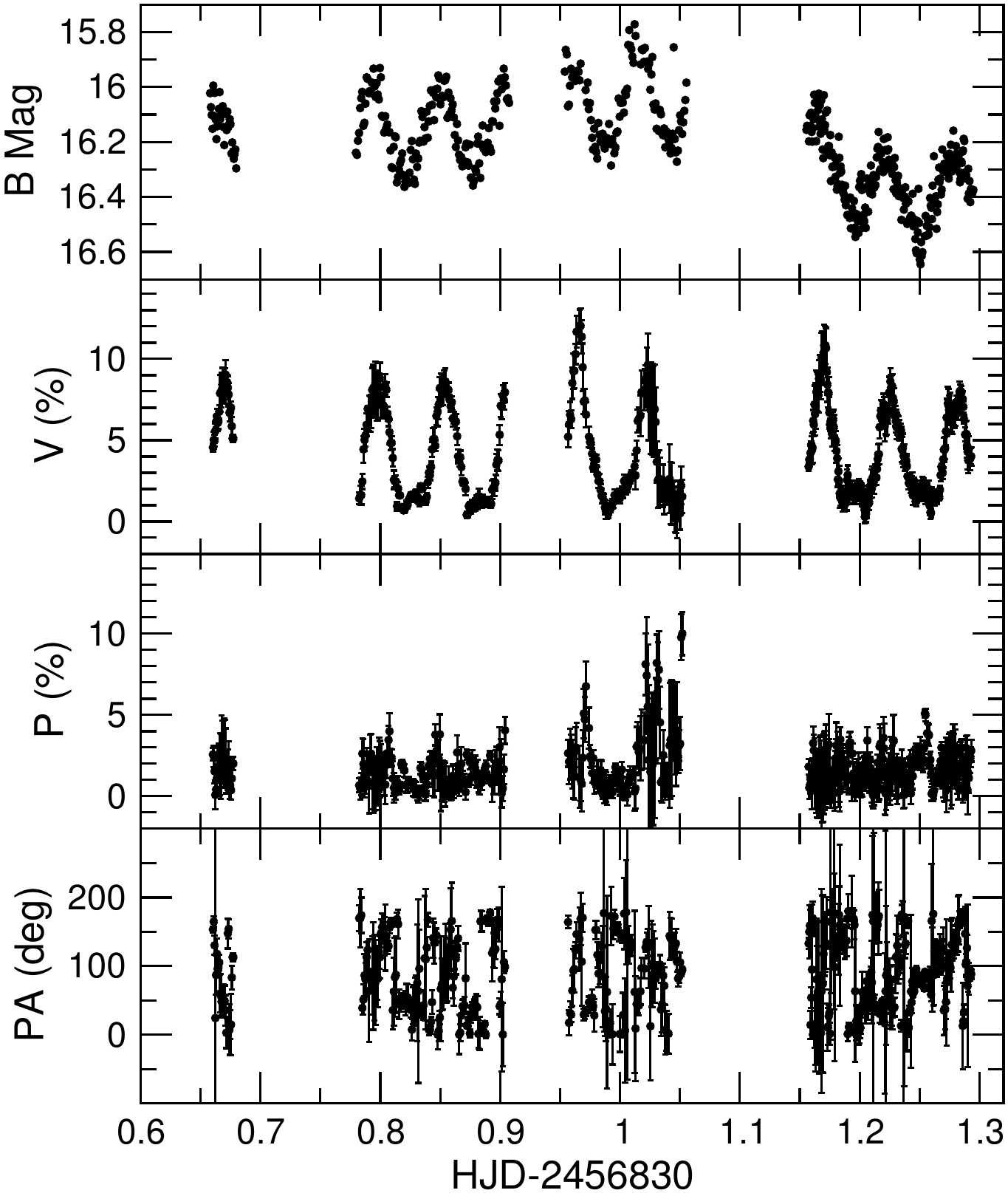}\par\medskip
 \includegraphics[width=\linewidth]{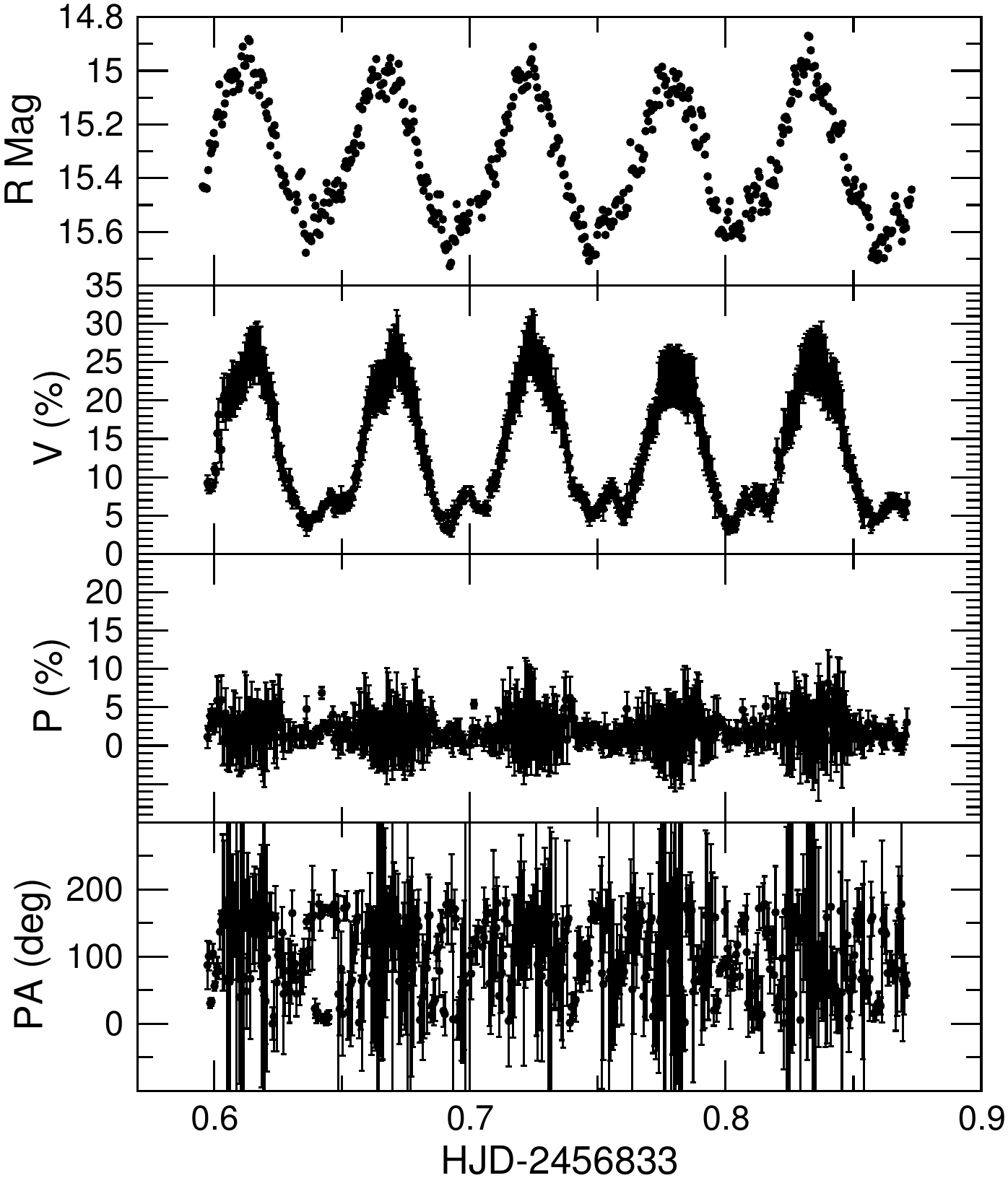}\par
 \includegraphics[width=\linewidth]{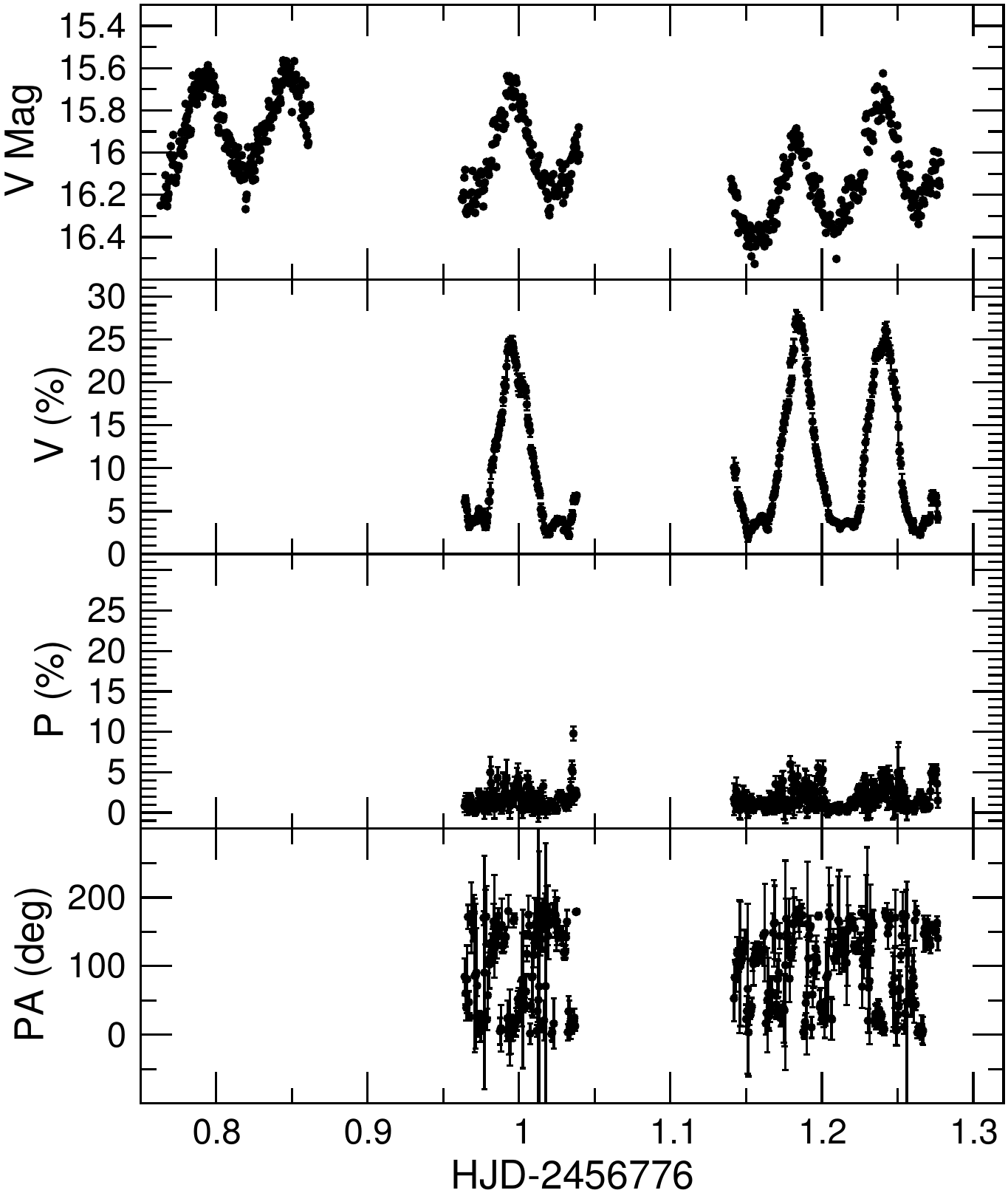}\par\medskip
 \includegraphics[width=\linewidth]{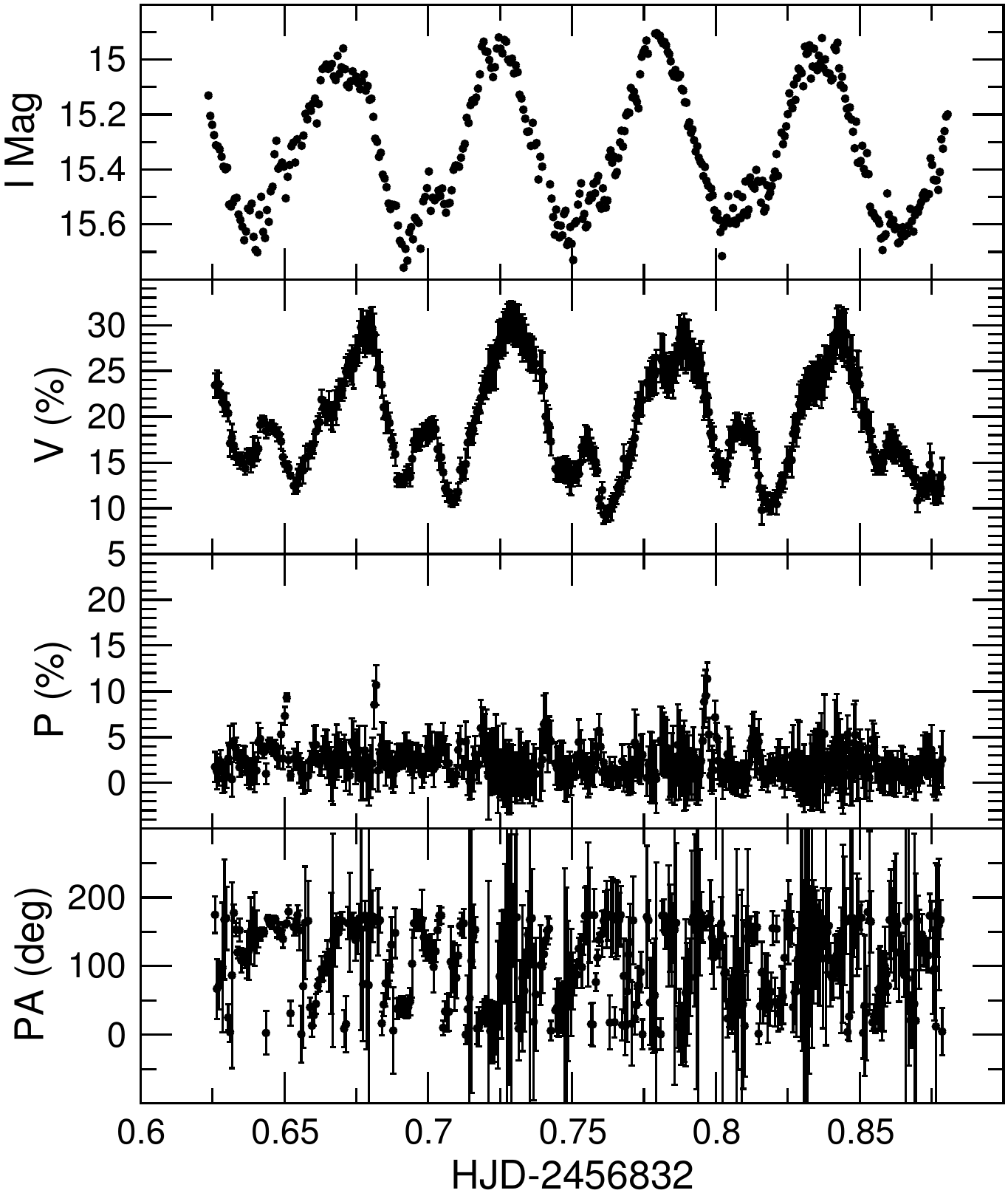}\par
\end{multicols}
 \caption{$B$, $V$, $R$ and $I$-bands light and polarization curves of \vvp. In each band the panels show, from top to bottom, the magnitude, circular polarization $V$, linear polarization $P$, and position angle $PA$ of the linear polarization. The abscissas of the first data set in each band are indicated in the axis label, while arbitrary constants were subtracted, when necessary, from the timings of the subsequent nights for clarity.}
 \label{fig:lightcurves}
\end{figure*}

The OPD datasets have higher temporal resolution and last up to 6.5 h each. The OPD magnitude light curves (upper panels of Fig.~\ref{fig:lightcurves}) show in all filters a quasi-sinusoidal modulation.  The amplitude has the maximum value in the $R$ band (around 1~mag). The smallest amplitude, of around 0.4~mag, occurs at the $B$ band. The light curves also show a daily fluctuation of the mean brightness of the system.

In order to search for periodicities in the OPD light curves, we applied the Lomb--Scargle (LS -- \citealt{1982ApJ...263..835S}) method to individual data sets, to filter-combined datasets, and also to the median-normalized complete dataset. We obtained a period of 0.05554~d (79.98~min) in all cases. The photometric ephemeris presented in equation~\ref{ephem_phot} was determined for the median normalized complete dataset, which has a 83-days length. The normalization by the median magnitude in each set was necessary because of the daily fluctuation in the average brightness of \vp and also to deal with the different magnitude levels in distinct filters. The zero phase of this ephemeris is defined as the minimum of the sinusoidal function fitted to the normalized photometric data. We additionally applied the LS method to the CRTS data, but could not recover the 0.05554~d period due to the poor CRTS data sampling. 

\begin{equation}
HJD_{\mathrm{min}} = 2456776.764(\pm6)+0.05554(\pm6) \times N 
\label{ephem_phot}
\end{equation}

\begin{figure}
\includegraphics[width=\columnwidth]{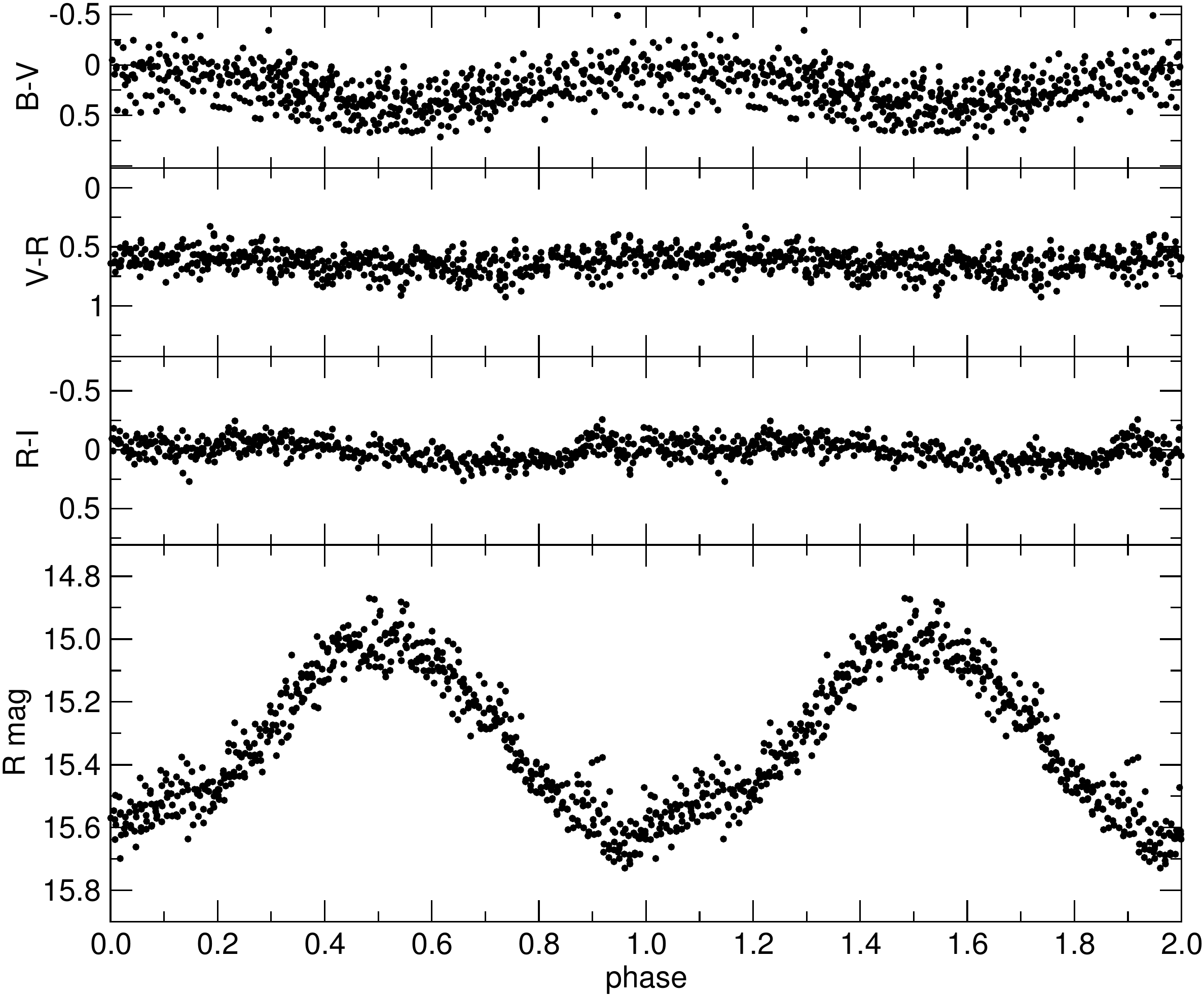}
\caption{Colour variations of OPD photometric data (upper panels) and phase diagram of the OPD data in $R$ band (bottom panel), folded using the photometric ephemeris from equation~\ref{ephem_phot}. The phase origin is at photometric minimum.}
\label{fig:photonorm}
\end{figure}

The phase diagram of the $R$ filter data and of the colour variations, both relative to the photometric ephemeris, are shown in Fig.~\ref{fig:photonorm}. The higher dispersion seen in the $B-V$ curve is caused by the slight difference in brightness states, associated to the daily fluctuation. From the $B-V$ curve one can note that \vp is bluer near phase zero, and redder half a cycle after that. The $V-R$ and $R-I$ curves are nearly flat. 

The long term \vp light curve (Fig.~\ref{fig:photoCRTS}) shows a change in the brightness level of approximately 3~mag amplitude, which is typical of polar CVs, like AM Her. Some polars present a clear separation between high- and low-brightness states, which is not the case of \vvp. Its magnitude varies smoothly with some rapid increases in flux during the periods of low brightness. Fig.~\ref{fig:photoCRTS} makes clear that our OPD photometric observations and the SOAR spectroscopic observations were obtained while \vp was at the high state. In polars, these states are understood as consequences of mass-transfer rate variations, either related to a temporary halting of mass-transfer due to the passing of magnetic star-spots in the inner Lagrangian point  \citep{1994ApJ...427..956L}, or to changes in the magnetic interaction between the WD and the mass-donor star \citep{2008A&A...481..433W}.

In order to search for a possible periodicity associated to the daily slow fluctuation present in the OPD data, we pre-whitened the data in all 4 bands with the 0.05554 d modulation and repeated the LS method to filter-combined and date-combined datasets. The resulting power spectra of the filter combined datasets show no consistent periodicity, while the $B$ plus $V$ data taken on 2014 July 20 display a possible, but uncertain, modulation at the 0.23~d (= 5.4~h) period. We do not claim this modulation is real, since it is neither persistent nor present in the other datasets. 

Fig.~\ref{fig:lightcurves} shows the \vp polarization as function of HJD in $B$, $V$, $R$, and $I$ bands. The high values of circular polarization unambiguously show that the \vp is a polar. The circular polarization increases from blue to red bands, reaching around 30 per cent in $R$ and $I$ bands. The linear polarization is very noisy (and hence also its position angle -- $PA$) and there is no evidence of a linear-polarization pulse. This fact together with the lack of change of the sign of the circular polarization indicates a configuration in which the magnetic field does not point to observer in any phase.

\section{Spectroscopic results}
\label{sec:spec_results}

\begin{figure}
 \includegraphics[width=\columnwidth]{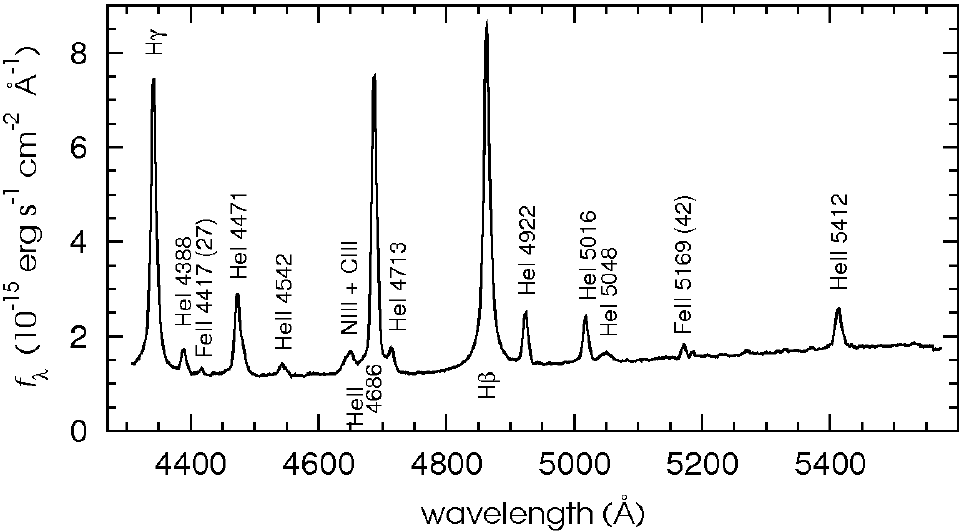}
  \caption{Average spectra of \vp obtained along 12 orbital cycles in June, 2014. The FWHM resolution is 1.6~\AA.}
\label{fig:ave_rspec}
\end{figure}

The average intermediate resolution spectrum of V348 Pav is presented in Fig.~\ref{fig:ave_rspec}.
A typical mCV emission line spectrum is shown with high \ion{He}{ii}/\ion{He}{i} line ratios. Low-ionization \ion{Fe}{ii} multiplet 27 and 42 lines are seen weak over the orbit averaged red continuum. No absorption features from the secondary star are detected. Synthetic photometry on a line-free band at 5225 \AA~ shows a quasi-sinusoidal continuum modulation consistent with the broadband photometry, with maximum at phase 0.50 of the spectroscopic ephemeris presented below, while the H~$\beta$ flux has a maximum around phase 0.35. 
This average spectrum is very similar to the spectrum of \vp obtained in 2012 \citep[see][]{2017AJ....153..144O}, both acquired at approximately the same high-brightness state (Fig.~\ref{fig:photoCRTS}). 
In the 2014 spectra, the \heii to H~$\beta$ line ratio reaches \ion{He}{ii}/H~$\beta$ $\sim0.9$. The difference between this value and the ratio \ion{He}{ii}/H~$\beta$ $\sim$ 0.25 obtained by \citet{1994AJ....107.2172D} 
is probably caused by varying mass accretion rate.

Multiple line components can be identified in the recombination lines. All \ion{He}{ii}, \ion{He}{i} and Balmer profiles show a strong narrow component with a sinusoidal radial velocity curve. This component is often recognised as the emission from the X-ray illuminated secondary. The velocities of the well-defined line peak in both Balmer, \hei~ and \heii profiles were measured by weighting the wavelength by a high power (8 to 16) of the line flux, yielding low-noise velocities of the line peak centroid.  A simple sinusoidal fit to H~$\beta$ narrow component was chosen to provide the following spectroscopic ephemeris for negative-to-positive gamma crossings (i.e. the inferior conjunction of the secondary):

\begin{equation}
HJD_{+/-} = 2456832.3064(\pm8)+0.05554(\pm5) \times N 
\label{ephem_spec}
\end{equation}

\noindent The spectroscopic period above is consistent within the uncertainties with the photometric period derived in previous section. 
The photometric phase coincides with the spectroscopic phase considering the epoch uncertainties.

\begin{figure}
  \includegraphics[width=\columnwidth]{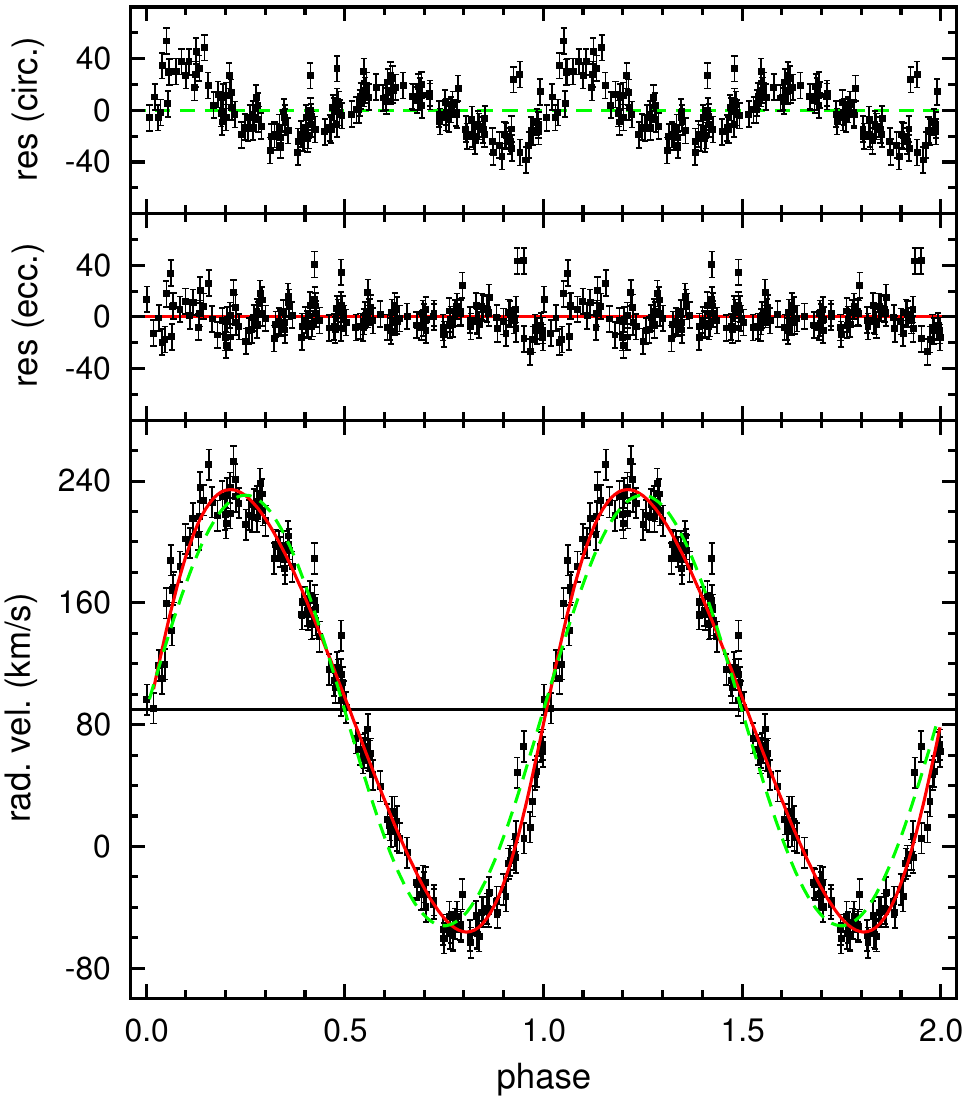}
  \caption{Radial velocity curve of the H~$\beta$ narrow component in \vp. The top two panels correspond to the residuals regarding a single sinusoidal fit (green) and eccentric fit (red). Both curves are phased according to the spectroscopic phase given by equation~\ref{ephem_spec}.}
\label{fig:rv_curves}
\end{figure}

Phase-folded radial velocity curves of H~$\beta$ narrow component are shown in Fig.~\ref{fig:rv_curves}. In this figure a pure sinusoidal fit is compared with an eccentric fit of H~$\beta$ velocities. Both circular and eccentric orbits were fitted to the radial velocity data using the RVFIT code by \citet{iglesias2015}. Eccentric fittings of H~$\beta$ and \heii lines produce statistically significant eccentric terms $e=0.14\pm0.02$ and $0.07\pm0.01$, respectively. Circular and eccentric fittings result in similar phasing and semi-amplitudes for those lines. Circularization time-scales in such short period binary can be as short as $10^3$~yr or less \citep{Hilditch2001}. While a better fit is provided by the addition of such a non-zero eccentricity, the coherent structures seen in the circular fit residuals, which were suppressed by the eccentric fit, have sub-resolution amplitudes ($\approx$~FWHM/4) and can be due to the contribution of unidentified non-photospheric components included in the peak centroid sampling. One may also speculate that this effect on measured radial velocities may be due to an orbit modulated aspect of the uneven companion illuminated surface and/or Balmer self-absorption by the stream, which change the effective emission position around the orbit and mimic an eccentric displacement of the companion. Therefore, the analysis of better spectral resolution data would be required to eventually claim the presence of an eccentric orbit in this system.

Despite the limited spectral resolution of the data, a  precise semi-amplitude $K_{\mathrm{em}}=141.4\pm1.5$ km s$^{-1}$ could be measured for the effective emission location at the secondary surface. The measured velocity reflects the Doppler shifted H~$\beta$ specific intensity summed over the visible secondary hemisphere and is not a direct measure of the secondary mass centre velocity, which could be used to derive the primary mass function. If one considers the illumination by an unshielded soft X-ray emission from the primary, a higher emissivity is expected near the inner Lagrangian point ($L1$) while a much larger area contributes near the terminator. The problem of correcting observed $K$ semi-amplitudes is essential in short period CVs given that the correction itself depends on trial masses. Therefore, each point at the mass diagram is calculated iteratively, assuming a Roche lobe-filling secondary. The $K-$correction factor $f\in[0.42,1.0]$ was first defined by \citet{1988ApJ...324..411W} as the light centre shift regarding the secondary barycentre in units of the secondary radius. Such a factor describes how the emissivity varies along the companion chromosphere, ranging from an even distribution over the illuminated lobe to an unrealistic point source at $L1$. 
The irradiation model presented by \citet{munoz-darias2005} suggests a $f$ value between $\sim$0.45 and $\sim$0.7 for \vvp. 
A factor $f=0.5$ would correct the observed radial velocity semi-amplitude $K_2$ to a range between 160 and 170~km~s$^{-1}$, depending on the system parameters.

Constraints on the orbital inclination are hardly found in non-eclipsing mCVs. The modelling of the cyclotron emission presented in Section \ref{sec:cyclops} suggests a low orbital inclination, which would be consistent with an intermediate mass white dwarf, considering a ZAMS Roche lobe-filling illuminated secondary. An inflated companion regarding the ZAMS radius \citep{Knigge11,cruz2018} translates in a lighter primary.

\subsection{Doppler tomography}
\label{sec:dopplermap}

\begin{figure}
  \includegraphics[width=\columnwidth]{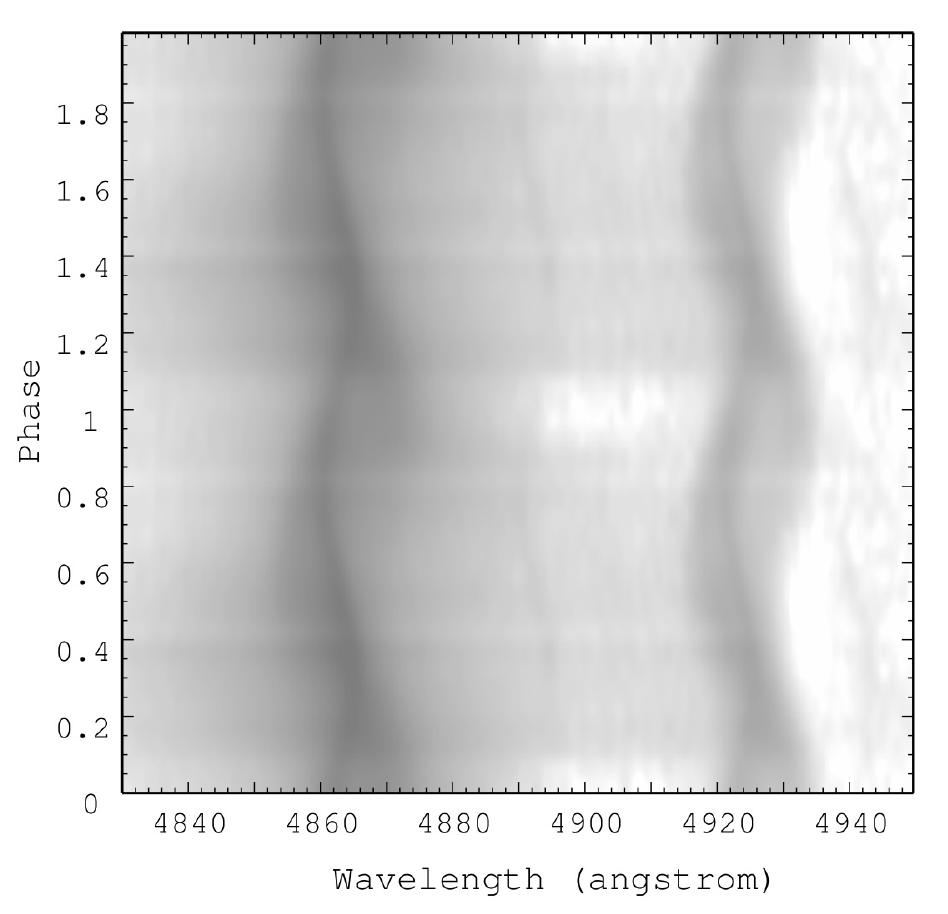}
  \caption{Trailed spectrogram of \vp phased according to equation~\ref{ephem_spec} in the H~$\beta$ plus \hei~ region. The phase resolution is 0.03 while the spectral resolution is 1.8 \AA~ (FWHM). A logarithmic grey-scale is shown to enhance the much fainter broad structures in the profiles.}
\label{fig:greenstein}
\end{figure}

Besides the sharp companion chromospheric emission, much fainter but extended broad components can be also seen in the trailed spectrogram of \vp (Fig.~\ref{fig:greenstein}). Those broad components in mCVs arise from the projection of complex intrinsic velocities in the accretion stream. Doppler imaging and analysis of the stream fluxes in the polar MR Ser suggests that a significant fraction of broad Balmer and \ion{He}{} lines are formed by photoionization of the ballistic stream and coupling region by the soft X-ray component in this system \citep{diaz2009}. In order to better understand the origin of all recombination line components, basic Doppler images of the emission lines were analysed. The filtered back-projection algorithm \citep{rosenfeld_kak(1982)} is used to compute the presented reconstructions of velocity maps in \vvp, with all input data provided by the observed line profiles. In that sense, the Doppler tomogram may be regarded as a convenient representation of the phase-resolved spectra where all the signal available along the series is linearly co-added. The usual coordinate system definition is displayed in our maps; the origin is at the binary rest, the x-axis is directed from the primary to the secondary star, while the y-axis is in the direction of the secondary orbital motion.

\begin{figure}
\includegraphics[width=\columnwidth]{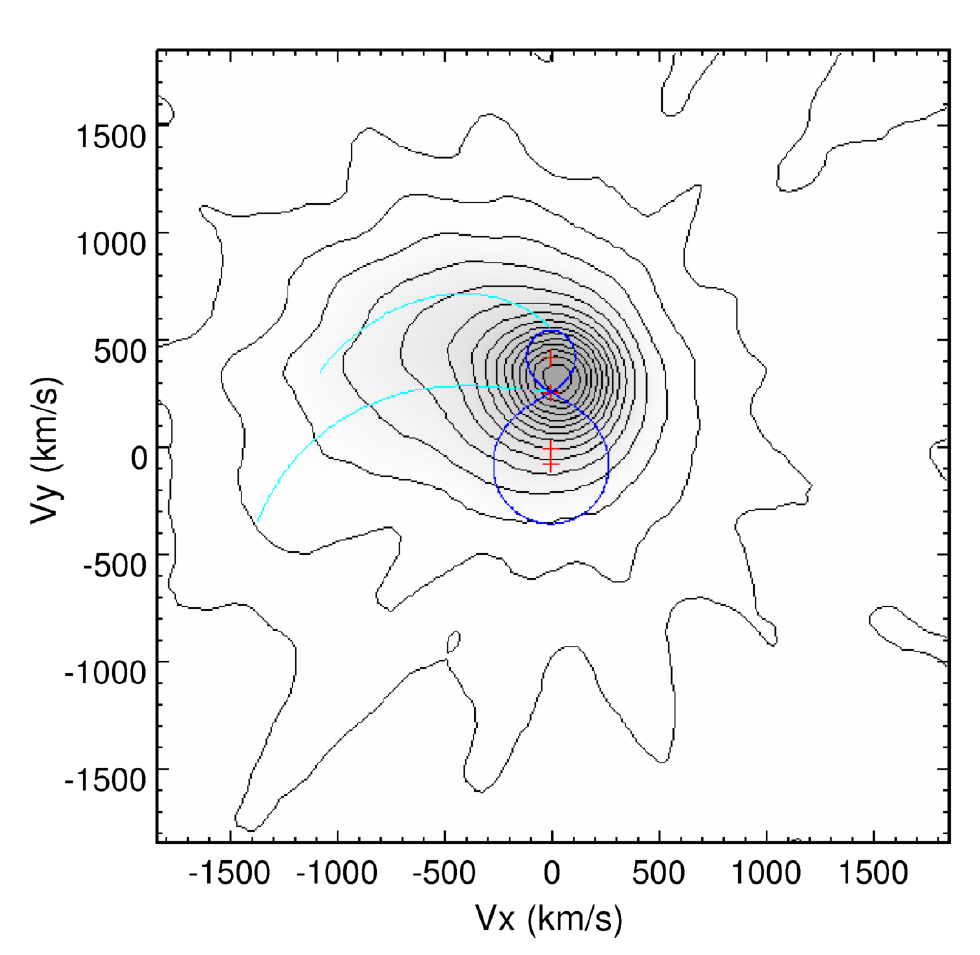}
  \caption{Doppler map of the H~$\beta$ emission in \vvp. The pluses correspond, from top to bottom, to the centre of mass of the secondary, the inner Lagrangian point L1, the binary mass centre, and the primary centre of mass. The upper curved line shows the corresponding keplerian velocities at the stream path inside the primary Roche lobe. The lower line represents the ballistic stream velocities. For visualisation purposes, the velocity scale corresponds to the deprojection from an orbital inclination of 25$^{\circ}$. A sample primary mass of 0.6~M$_{\sun}$ was assumed for plotting the binary components. The 16 contour levels are equally spaced, from 0.0 to the maximum observed flux density of $4.2\times10^{-19}$erg~cm$^{-2}$~s$^{-1}$(km~s$^{-1}$)$^{-2}$. The map resolution is 260~km~s$^{-1}$~(FWHM).}
\label{fig:dopplermaphb}
\end{figure}

Doppler reconstructions of H~$\beta$ and \heii emission were calculated using all available medium-resolution spectra. In both maps the companion chromospheric emission is the strongest structure seen (Fig.~\ref{fig:dopplermaphb}). This feature has a symmetric core with a FWHM of 270 km s$^{-1}$ and 210 km s$^{-1}$, which accounts for 35 and 31 per cent of the total system emission in H~$\beta$ and \hheii, respectively. A slight asymmetric extension of the companion illumination can be seen towards the expected ballistic stream at lower emissivities. Most of the high-velocity extended emission is located in the $V_{x}, V_{y} < 0$ quadrant which may suggest an origin in the coupling region or in the accretion column itself. 

\begin{figure} 
\includegraphics[width=\columnwidth]{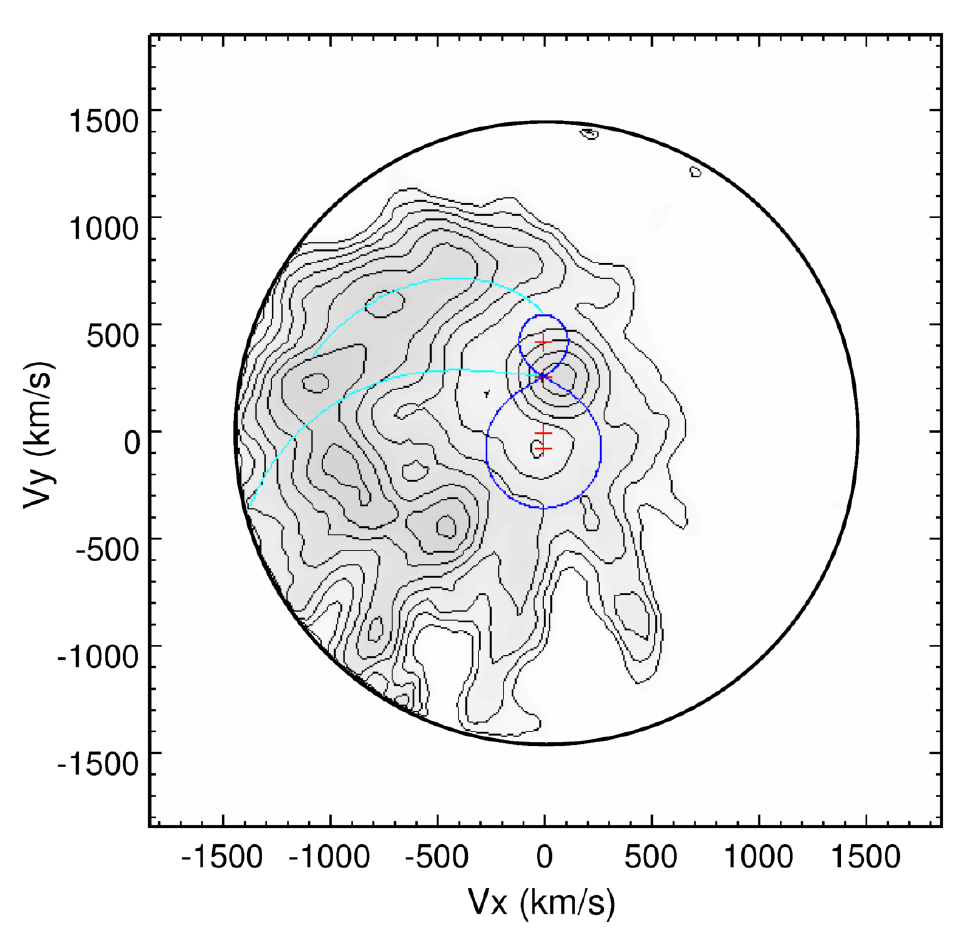}
  \caption{Same as Fig.~\ref{fig:dopplermaphb} for the \heii to H~$\beta$ emissivity ratio. Contour lines are equally spaced by 0.1, from 0.6 to 1.2. The reconstruction noise is amplified at faint H~$\beta$ regions. Beyond the circular subraster shown it raises above 30 per cent.}
\label{fig:doppratio}
\end{figure}

The \heii map is compared with Balmer emission by their emissivity ratio in  Fig.~\ref{fig:doppratio}. A stronger \ion{He}{ii} emission is seen near the inner Lagrangian point indicating a high-ionization state of the illuminated companion atmosphere. Although the observation and reconstruction noise is amplified in rational maps, an excess of \ion{He}{ii} emission relative to H~$\beta$ is clearly present for $V_{x} < 0$. Depending on the dipole longitude (see next section) an enhanced high-ionization recombination emission is expected at higher densities in the accretion column with negative $V_{x}$. However, those features with high intrinsic velocities, often out of the orbital plane, are subject to a significant blurring by a ring-shaped convolution kernel. This effect also known as gamma-smearing arises in the attempt to map 3-D structures without extra data or model input to the Doppler imaging method. As a result, the column emission may spread over a wider area and may not be identified as such in the Doppler reconstructions of mCVs (e.g., \citealt{diaz1994}).

\section{Cyclotron emission modelling}
\label{sec:cyclops}

Cyclotron radiation from the post-shock region (PSR) near the WD surface produces most of continuum emission in optical and infrared bands in polars and accounts for their variability in these wavelengths \citep[e.g.,][]{cropper1990}. This emission is a function of the temperature, density, and magnetic field distributions in the PSR as well as of its geometry and inclination. 
\citet{aizu1973} is one of the seminal papers regarding the hydrodynamics of accretion flows in WDs.
Recent examples of studies on the cyclotron emission in non-homogeneous PSR are \citet{wf1988}, \citet{ww1990}, \citet{ww1992}, \citet{potter2004}, and \citet{sarty2008}.  

We performed the cyclotron emission modelling of \vp using the \cyc code. \cyc considers a 3-D non-homogeneous PSR based in a  bipolar magnetic field. The emitting region is represented by a 3D grid of points of different densities, temperatures and magnetic fields. In each rotational phase, the radiative transfer is done for a bunch of line of sights that represents the projection of the PSR in the plane of the sky. In each line of sight, we consider a minimum number of points in order to appropriately take into account the shock structure. More details about the radiative transfer treatment and the code geometry can be found in \citet{costa2009} and \citet{silva2013}.

The present results were obtained with an updated version of the \cyc code. It solves the stationary one-dimensional hydro-thermodynamic equations describing the accreting plasma, where bremsstrahlung and cyclotron radiative processes play a role
in cooling the gas from the shock till the WD surface. This upgraded version allows accurate modelling of the PSR and a summary of the main changes is provided in what follows, but more details can be found in Belloni et al. (in preparation). We have taken into account the WD gravitational potential \citep[e.g.][]{cropper1999} and considered equipartition between ions and electrons, which is a reasonable assumption for most polars \citep[e.g.][]{vanboxsom2018}. We have adopted a quasi-dipolar geometry (i.e. a cubic cross-section variation) following \citet{hayashi2014}, allowed the WD magnetic field to decay as the distance from the WD surface increases \citep{Canalle_2005} and taken into account that the coupling region is not at infinite, as described in \citet{Suleimanov_2016}.
The plasma cools down from the shock until the WD surface due to effects of both bremsstrahlung and cyclotron radiative processes. Bremsstrahlung cooling and the relative importance between the two processes are described as power-law functions following \citet{vanboxsom2018}, and cyclotron cooling is described following \citet{Canalle_2005}. We would like to stress that PSR profiles obtained with such an updated version of the \cyc code are consistent with previous works \citep[e.g.][]{cropper1999,yuasa2010,
hayashi2014,vanboxsom2018}, provided the assumptions are the same in the modelling.

The \cyc models for \vp are calculated for the $B$, $V$, $R$, and $I$ data, which were folded in 40 phase bins relative to the photometric ephemeris (Equation \ref{ephem_phot}). Due to the brightness changes of \vvp, which possibly indicate changes in the PSR, we performed the modelling using the data taken in 2014 June, which cover 4 consecutive nights. The current version of \cyc allows the use various frequencies to represent a band. We verified that for the $B$, $V$, and $R$ bands a single frequency is enough to a good representation of the band emission, while five frequencies are needed for the $I$ band. As the linear polarization of this object is small and very noisy, only the total flux and the circular polarized flux were considered in the search for the best-fitting model. For the same reason, the $PA$ of the linear polarization was also ignored in the analysis of the figure of merit of the fits. 

Our modelling procedure aims to reproduce the relative fluxes among bands so, as a consequence, the best fit model also represents the \vp spectral energy distribution (SED) shape in the observed optical bands. This is an important constraint to the models because the cyclotron emission is highly modulated in wavelength due to its harmonics distribution. 

The search of the best model for \vp has been done in two steps. Initially, we performed a search in a very open range of parameters using the {\sc pikaia} code \citep{pikaia}, which is based on a genetic algorithm. In this search, we run almost $10^{5}$ models. Then, the best models are refined using the {\sc amoeba} code \citep{num_recipes}. The figure of merit of a model fit is the $\chi^2$ using the same weight for all points and bands, but we adopt a weight 10 times larger for the circularly polarized flux relative to the total flux.

\begin{figure*}
\begin{multicols}{2}
 \includegraphics[trim=0mm 0mm 12mm 10mm,clip,width=\linewidth]{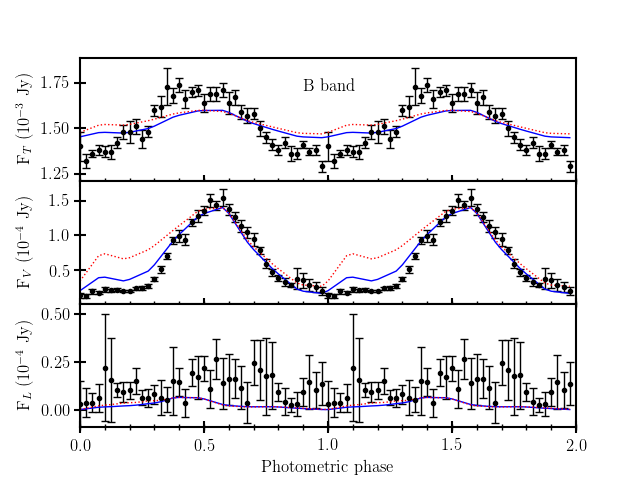}\par
 \includegraphics[trim=0mm 0mm 12mm 10mm,clip,width=\linewidth]{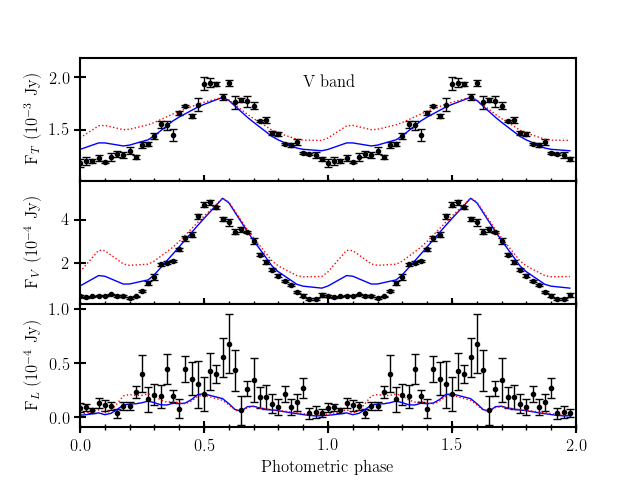}\par
 \includegraphics[trim=0mm 0mm 12mm 10mm,clip,width=\linewidth]{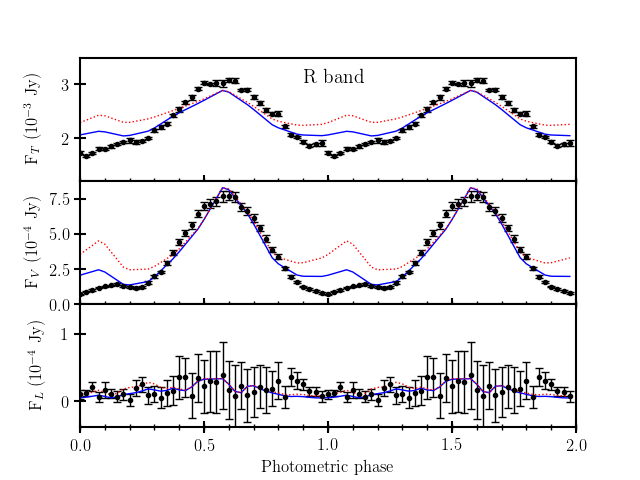}\par
 \includegraphics[trim=0mm 0mm 12mm 10mm,clip,width=\linewidth]{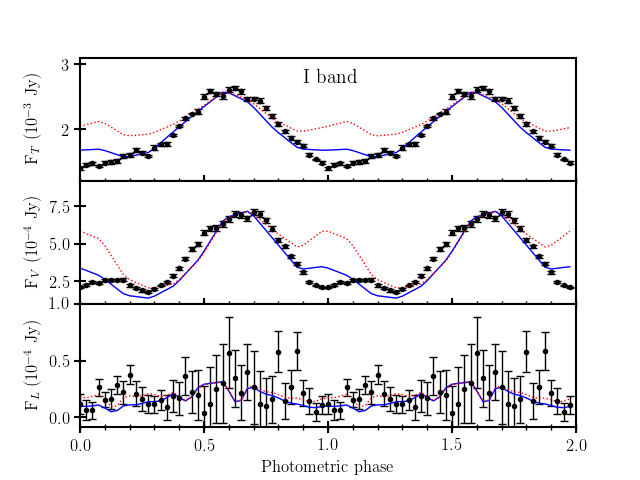}\par
\end{multicols}
 \caption{{\sc cyclops} best model (blue, solid line) over observed data (black errorbars) for $B$, $V$, $R$ and $I$ filters, as a function of the photometric phase (Equation \ref{ephem_phot}). The red dotted line represents the normalized model with no internal scattering in the pre-shock column. For each band, from the upper to the bottom panels, are shown the total (polarized and unpolarized) flux ($F_{\mathrm{T}}$), the circularly polarized flux ($F_{\mathrm{V}}$) and the linearly polarized flux ($F_{\mathrm{P}}$). }
 \label{fig:cyclops_models}
\end{figure*}

Fig.~\ref{fig:cyclops_models} (blue solid lines) presents the best fit of our {\sc cyclops} modelling. 
Consistent with the data, the best model presents negligible values of linear polarization, despite disregarding  the linear polarization as a constraint.
The parameters of this model are shown in Table~\ref{tab_cyclops} and discussed in the next paragraphs.

\begin{table}
\caption{Physical and geometrical properties of V348 Pav estimated from {\sc cyclops} modelling. 
} 
\label{tab_cyclops}
\begin{center}
\begin{tabular}{lc} 
\hline
\cyc input parameters  & Fitted values \\
\hline 
White dwarf mass, $M_{WD}$ & 0.85 M$_{\sun}$ \\
Mass accretion rate, $\dot{M}$ & $10^{-11}$ M$_{\sun}$ yr$^{-1}$ \\
Magnetic field intensity in the pole, $B_{\mathrm{pole}}$  &  28$\times 10^{6}$~G \\
Latitude of the magnetic axis, $B_{\mathrm{lat}}$ &  67$\degr$  \\
Longitude of the magnetic axis, $B_{\mathrm{long}}$  &  31$\degr$ \\
Orbital inclination, $i$ & 25$\degr$  \\
Colatitude of the PSR, $\beta$ &  10$\degr$  \\
Azimuthal semi-extension of the threading region  &  54$\degr$ \\
Radial semi-extension of the threading region &  0.7 \\
\hline
\cyc model associated and resulting quantities & Values \\
\hline
Specific accretion rate, $a$, in the WD surface &  0.0065~g\,cm$^{-2}$\,s$^{-1}$\\
Height of the PSR, $h$ &  0.016 R$_{\mathrm{WD}}$ \\
Shock electronic temperature
&  37~keV \\
Shock electronic number density
& $4\times 10^{13}$ cm$^{-3}$\\
Shock electronic mass density & $4 \times 10^{-11}$ g~cm$^{-3}$\\
Size in latitude of the PSR footprint & 19\degr \\
Footprint area relative to the WD surface & 0.017 \\
Distance of the threading region to the WD centre & 12 $R_{WD}$ \\
Magnetic field in the PSR & 24 -- 27 MG\\
Phase shift applied to the model, $\delta_{\mathrm{phase}}$  & -0.317 \\
Added flux in $B$ band & 0.0014 Jy \\
Added flux in $V$ band & 0.0012 Jy \\
Added flux in $R$ band & 0.0018 Jy \\
Added flux in $I$ band & 0.0012 Jy \\
\hline 
\end{tabular} 
\end{center}
\end{table}

The pole magnetic field and WD mass provided by the model are 28~MG and 0.85~$M_\odot$, respectively, which are typical of polars \citep{lilia2015,2011A&A...536A..42Z}. 

\begin{figure}
\centering
\includegraphics[trim=20mm 0mm 0mm 0mm,clip,width=\columnwidth]{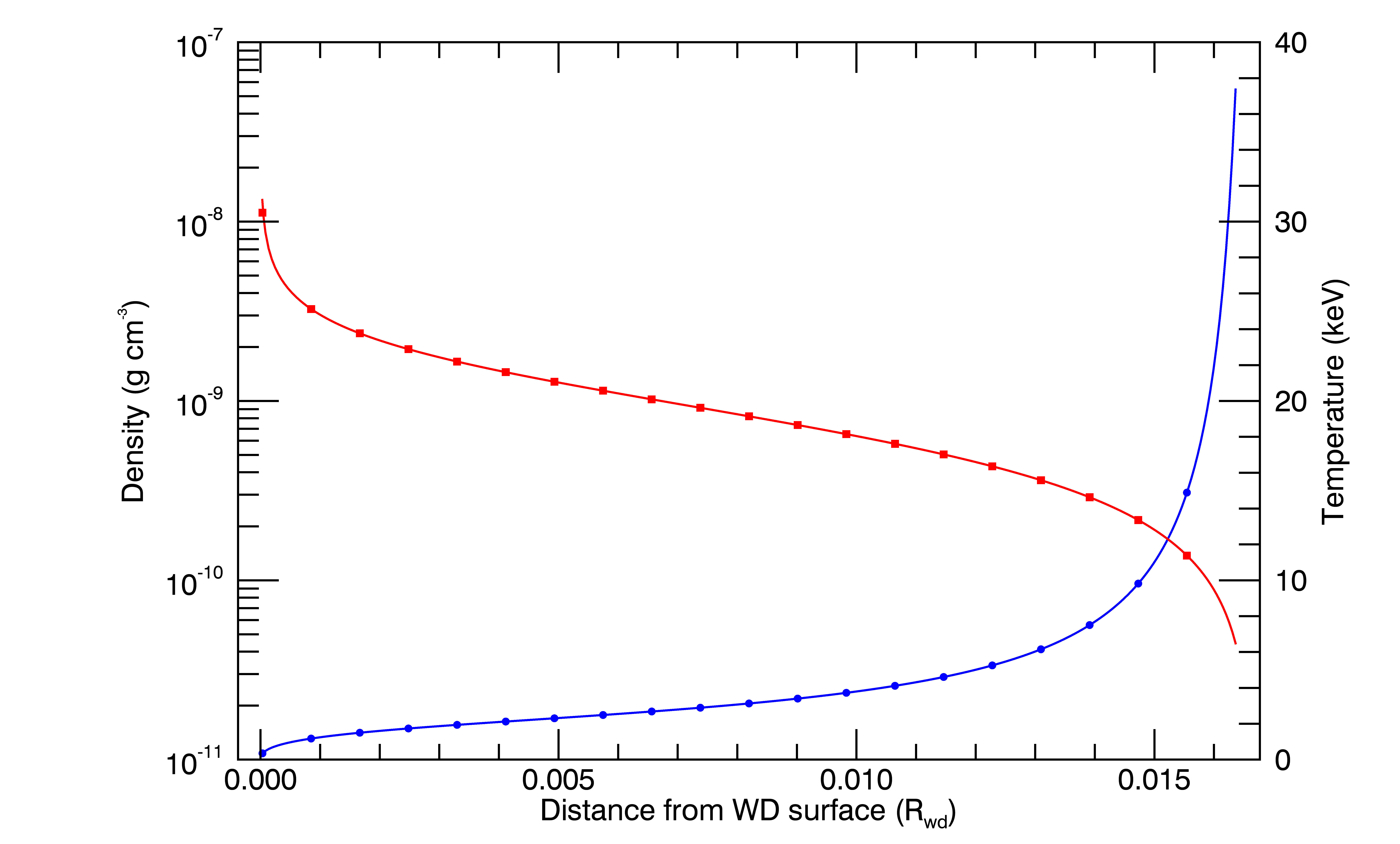}
\caption{Density (red squares) and temperature (blue circles) profiles of the PSR in the best-fitting model of \vvp.} 
\label{fig:shock_structure}
\end{figure}

The PSR density and temperature profiles of the best-fitting model for \vp are shown in Fig.~\ref{fig:shock_structure}. 
Notice that the density smoothly increases through the PSR, from the shock until the WD surface, but rapidly increases near the PSR bottom. Such a feature is consistent with previous works and is a consequence of the sudden decrease of the plasma velocity while reaching the WD surface \citep[e.g.][]{hayashi2014,vanboxsom2018}. Additionally, at the PSR top, it significantly differs from a power-law distribution, since in our modelling a dipole-like geometry is assumed as well as a magnetospheric boundary and non-negligible cyclotron cooling. Unlike the density profile, the temperature profile more strongly depends on the 
model parameters in particular, on the specific accretion rate and on the WD magnetic field at the PSR bottom (see Belloni et al., in preparation, for more details). The shock temperature is $\approx37$~keV, which is smaller than expected from models in most previous works for a WD mass of
$\approx0.85$~M$_{\sun}$ and a specific accretion rate at the WD surface of $\approx0.0065$~g\,s$^{-1}$\,cm$^{-2}$. This is because we assumed that the coupling between the accretion stream from the donor and the WD magnetic field lines occurs at a finite distance
\citep[][]{Suleimanov_2016}, which leads to smaller velocities at the shock and, in turn, smaller temperatures. In addition, the temperature profile is flatter than expected in the case of bremsstrahlung-dominated flows. Indeed, the WD magnetic field enhances the
cooling, which leads the temperature to considerably fall off close to the PSR top. The non-negligible cyclotron radiation also makes
the PSR height smaller for the same reason.

The total accretion luminosity, $L_{\mathrm{acc}}$, can be estimated by the PSR emission, which is the sum of the cyclotron (optical) and X-rays luminosities, $L_{\mathrm{cyc}}$ and $L_{\mathrm{X}}$, respectively. In X-rays, the total emission is expected to come from the PSR. But, in optical wavelengths, other components can be present. Considering the basic parameters of the WD and the mass-loss rate, $\dot{M}$, we can estimate $L_{\mathrm{acc}}$ as

\begin{equation}
\centering
L_{\mathrm{acc}}  =  L_{\mathrm{cyc}} + L_{\mathrm{X}} = \frac{G   M_{\mathrm{WD}}}{R_{\mathrm{WD}}}\dot{M}.  
\label{eq_lacc}
\end{equation}

Using the above relation, we can verify if $\dot{M}$ obtained by our fitting is consistent with the \vp luminosity estimated from observations.
In the following estimates, we use the \vp distance from \textit{Gaia} DR2, which is 154.4 $\pm$ 1.8~pc \citep{Gaia2016,Brown2018}. 

A possible \vp X-ray counterpart is detected 5.7 arcsec away from its optical position by \textit{XMM-Newton} in the slew data mode, with 1.667~counts~s$^{-1}$ total band count rate (0.2~--~12~keV), which corresponds to a flux of $(5.3\pm1.9)\times10^{-12}$~erg~cm$^{-2}$~s$^{-1}$. This detection is registered as XMMSL1~J195647.7-603428 in the \textit{XMM-Newton} Slew Survey Clean Source Catalog \citep{2008A&A...480..611S}. 
Using the GAIA distance and the X-ray flux, the X-ray luminosity of the system is 1.50 $ \times 10^{31}~$erg~s$^{-1}$. 

The optical modelling of \vp assumes an additional component to the cyclotron emission, whose fluxes are shown in Table~\ref{tab_cyclops}. Using the observed fluxes in \textit{B, V, R}, and \textit{I} bands and removing the flux that does not have a cyclotron origin, we estimate that $L_{\mathrm{cyc}}$ is around $5.14~\times~10^{30}$~erg~s$^{-1}$. So we obtained $L_{\mathrm{acc}}~=~2.01~\times~10^{31}$~erg~s$^{-1}$. Equation~\ref{eq_lacc} implies in $\dot{M}$ equal to 1.2$~\times~10^{14}$g~s$^{-1}$, which is equivalent to 2$~\times~10^{-12}$~M$_{\odot}$~yr$^{-1}$. This is a lower limit 
to $\dot{M}$, because the cyclotron SED of our model extends out of the observed region and also because in high-energies the emission could come from other spectral regions not included in the above estimate.

As we aim to fit the SED, we implemented the application of the interstellar reddening to PSR emission in the code. The code input is the Hydrogen column density, N(H), which is converted to extinction considering $R$ = 3.1 and the extinction law of \citet{ccm89}.
In the direction of \vvp, $N(\mathrm{H})= 4.14 \times 10^{20}$~cm$^{-2}$~\footnote{https://heasarc.nasa.gov/cgi-bin/Tools/w3nh/w3nh.pl}, which is very small and consequently irrelevant to the fit.

\begin{figure*}
\centering
\includegraphics[trim=40mm 150mm 10mm 50mm,clip,width=\textwidth]{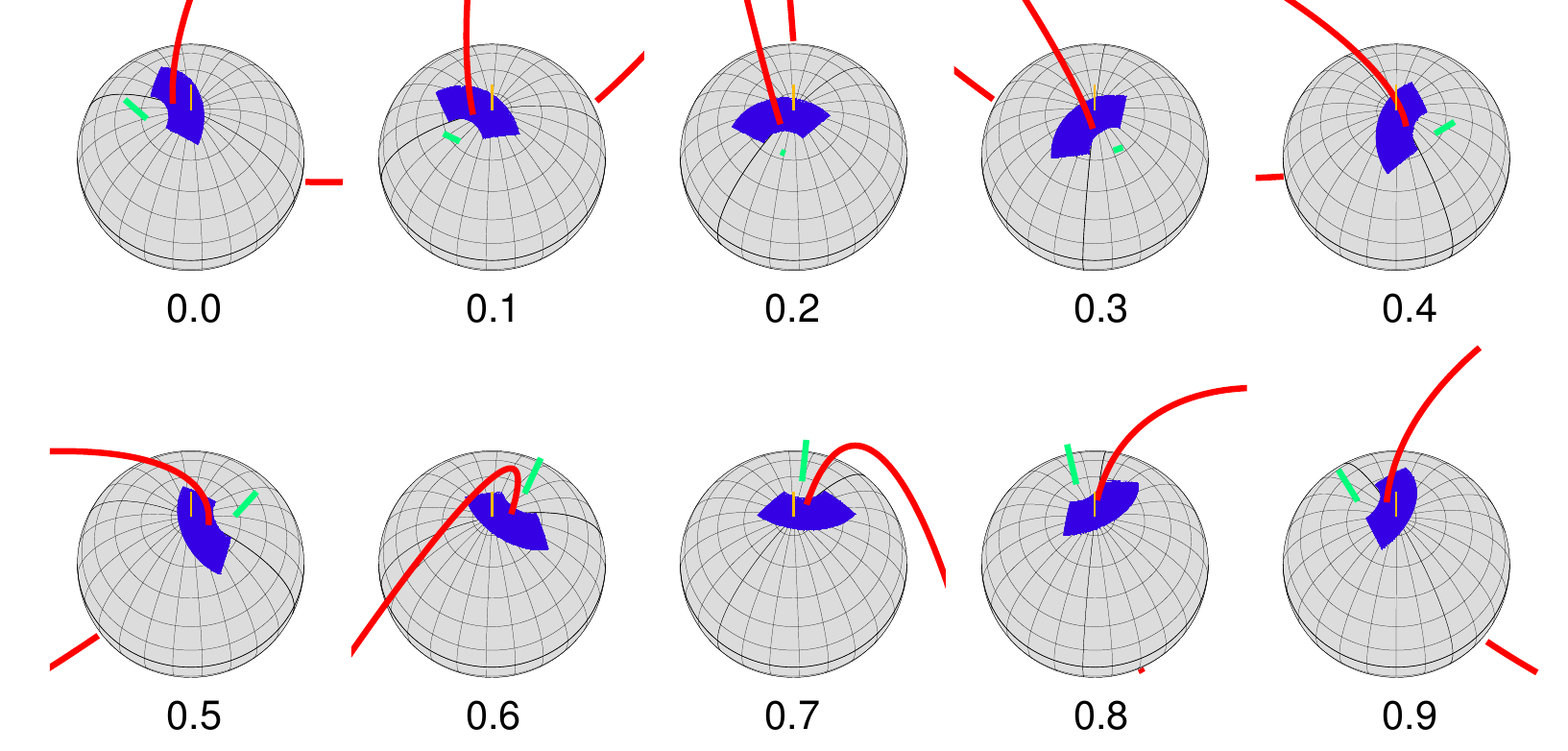}
\caption{Accretion geometry of \vp as seen by the observer in different phases. The 3D PSR on the WD surface is shown in blue. The reference magnetic field line that passes through the central point of the PSR is shown in red. The $B$ axis is show by the green straight line and the WD rotation axis, by the orange line.} 
\label{fig:accretion_geom}
\end{figure*}

Figure \ref{fig:accretion_geom} shows how the observer sees the WD and the PSR along the rotation cycle in the proposed model. The PSR contains the rotation pole of WD and is never obscured by the WD, i.e. there is no total or partial self-eclipse of the PSR. Hence all the flux modulation comes from the anisotropy of the cyclotronic emission. This figure also shows that the PSR is seen by the observer from above.

The entire accretion column follows a geometry defined by the magnetic field lines (see fig. 1 of \citealt{costa2009}).
In \vvp, the accretion column backbone is the red line of Fig.~\ref{fig:accretion_geom}.
Consequently, the pre-shock region is, in all phases, between the PSR and the observer. 
Many polars have their high-energy light curves explained by absorption of the PSR emission by material in the accretion column. Consistent with this scenario, the \cyc code includes the scattering of the optical PSR emission out of the optical path by electrons in the accretion column. 
The red dotted lines in Fig.~\ref{fig:cyclops_models} represents the best model, but with no pre-shock Thomson scattering taken into account. We see that this scattering is in fact important and can be relevant in shaping the observed flux and polarization curves of polars depending on the system geometry.

The secondary is irradiated by the PSR emission and also by the soft X-ray emission arising from the heated WD surface near the accretion footprint. The location of the PSR, near the rotational pole, is not the more appropriate for the secondary illumination, but part of this high-energy combined emission, which is isotropic, continuously reaches the secondary. 
Being a polar, the system is completely synchronised, so the relative position between the PSR and the secondary does not change along the WD rotation/orbital cycle.
Because the phasing of the photometric and spectroscopic ephemerides are the same within the errors (see previous sections), we can assume that at photometric zero phase the system is approximately at inferior conjunction. Fig.~\ref{fig:accretion_geom} shows that at this phase the accretion column does not intercept the optical path to the secondary direction (perpendicular to the WD rotation axis). 
Therefore, the emission from the PSR and from the heated WD region in the direction of the secondary is not obscured by the accretion column. 
Additionally, the pre-shock column can also redirect the high-energy emission, by scattering, to the secondary, also contributing to its heating. 
All these facts and the small distance between the stars in \vp are consistent with the narrow component of the Balmer lines being produced on the irradiated secondary.

We finish this section by discussing a caveat related to one assumption in this work. Ions and electrons are assumed here to be
in equipartition throughout the PSR, which is a reasonable approximation for most polars. Indeed, equipartition between ions and
electrons is a plausible assumption provided that the electron-ion equipartition time-scale is much shorter than the cooling
time-scale. However, when either the specific accretion rate is very small and/or the WD magnetic field is sufficiently strong, such an assumption is no longer valid \citep[e.g.][]{sarty2008}. That said, even though the \cyc code is not designed for investigations of polars with extreme specific accretion rates and/or WD magnetic fields, it is still a good computing machinery, since it is a reasonable first-order approximation in those cases. We note that involving more realistic modelling, in which non-equipartition between ions and electrons is considered, would be useful to further test the results achieved in this work.

\section{Conclusions}
\label{sec:concl}

In this work we have confirmed the nature of \vp as a polar, in view of its high circular polarization and intense \heii spectral emission line. We have presented the \vp light and polarization curves in the \textit{B, V, R,} and \textit{I} bands. The CRTS \vp observations have shown a variation of 2.5~mag in the average brightness of the system in a time scale of 9 years. All these features are also typical of polars. 

We have obtained spectroscopic time series spanning around 12~orbital cycles, deriving an H$~\beta$ radial-velocity curve semi-amplitude of $141.4\pm1.5$ km s$^{-1}$.

We determined the orbital period of \vp as $79.98\pm0.08$~min, both from radial velocities of the H~$\beta$ narrow component and from photometric measurements. This determination places \vp very close to the orbital period minimum (P$_{\mathrm{min}}$) of CVs, defined as the point where the evolution of the period changes direction due to the inversion of the response of the donor star radius to mass loss. \citet{Knigge11} sets P$_{\mathrm{min}}$ at $81.8\pm0.9$ min, based on the observed M-R relationship of donor stars. With measurements of system parameters for 15 eclipsing CVs, \citet{2019MNRAS.486.5535M} found a refined, lower value of P$_{\mathrm{min}}=79.6\pm0.2$ min. The \vp period leads to a secondary mass upper limit of 0.09~$M_\odot$ \citep{Knigge11}.

The Doppler tomography shows that the X-ray illuminated secondary star is the dominant source of the narrow spectral emission features. Broad asymmetrical emission, associated to the coupling region or to the accretion column, is also present in the data.
The radial velocity curves of the H~$\beta$ and \heii narrow components are better fit using eccentric terms. However, our data do not have enough spectral resolution to constrain its origin, i. e., whether it is due to the emission distribution in the binary system or rather unlikely orbital.

We modelled the behaviour of the Stokes parameters as a function of the WD rotation phase using a new version of the \cyc code. That version solves the 3D radiative transfer in the post-shock region, whose physical properties have been obtained from an one-dimensional solution of the hydro-thermodynamics. The best-fit model provides $M_{WD} = 0.85~M_\odot$, $B = 28~MG$ in the magnetic pole, and $\dot{M} = 10^{-11}$ M$_{\sun}$ yr$^{-1}$. This mass-accretion rate is consistent with the lower limit of the accretion luminosity estimated from X-rays and optical fluxes and GAIA distance. The system is seen by a low orbital
inclination and the post-shock region is located in the WD rotation pole. Hence, the base of the accretion column is never self-eclipsed along the WD rotation.

Due to the low inclination, any dynamical/spectroscopic determination of the white dwarf mass would be very uncertain. Therefore, our WD mass estimate is obtained from the {\sc cyclops} modelling only. Nevertheless, its value is consistent with the orbital inclination and the measured radial velocities. Higher spectral-resolution time-series data combined with IR light curves would be valuable to constrain the WD mass and elucidate the origin of the asymmetric radial velocity curve.

\section*{Acknowledgements}

CVR thanks Joaquim Costa for fruitful discussions about the \cyc code.
This paper is based on observations obtained at the Southern Astrophysical Research (SOAR) telescope, which is a joint project of the Minist\'{e}rio da Ci\^{e}ncia, Tecnologia, Inova\c{c}\~{o}es e Comunica\c{c}\~{o}es (MCTIC) do Brasil, the U.S. National Optical Astronomy Observatory (NOAO), the University of North Carolina at Chapel Hill (UNC), and Michigan State University (MSU), and on observations obtained at the Observatório do Pico dos Dias/LNA. ASO acknowledges São Paulo Research Foundation (FAPESP) for financial support under grant \#2017/20309-7. MSP thanks CAPES for financial support under grant \#88887.153742/2017-00. MPD thanks CNPq funding under grant \#305657. 
CVR is grateful to the support of {\it Conselho Nacional de Desenvolvimento Cient\'ifico e Tecnol\'ogico} - CNPq (\#303444/2018-5) and grant \#2013/26258-4, Funda\c c\~ao de Amparo \`a Pesquisa do Estado de S\~ao Paulo (FAPESP).
DB was supported by the grants \#2017/14289-3 and \#2018/23562-8, S\~ao Paulo Research Foundation (FAPESP). 
The authors thank to MCTIC/FINEP (CT-INFRA grant 0112052700) and the Embrace Space Weather Program for the computing facilities at INPE.
The CSS survey is funded by the National Aeronautics and Space Administration under Grant No. NNG05GF22G issued through the Science Mission Directorate Near-Earth Objects Observations Program.  The CRTS survey is supported by the U.S. National Science Foundation under grants AST-0909182 and AST-1313422.
This research has made use of data, software and/or web tools obtained from the High Energy Astrophysics Science Archive Research Center (HEASARC), a service of the Astrophysics Science Division at NASA/GSFC and of the Smithsonian Astrophysical Observatory's High Energy Astrophysics Division.
This work has made use of data from the European Space Agency (ESA) mission
{\it Gaia} (\url{https://www.cosmos.esa.int/gaia}), processed by the {\it Gaia}
Data Processing and Analysis Consortium (DPAC,
\url{https://www.cosmos.esa.int/web/gaia/dpac/consortium}). Funding for the DPAC
has been provided by national institutions, in particular the institutions
participating in the {\it Gaia} Multilateral Agreement.





\begin{thebibliography}{99}

\bibitem[\protect\citeauthoryear{Aizu}{1973}]{aizu1973} Aizu K., 1973, PThPh, 49, 1184
\bibitem[\protect\citeauthoryear{Canalle, Saxton, Wu, Cropper \& Ramsay}{2005}]{Canalle_2005} Canalle J.~B.~G., Saxton C.~J., Wu K., Cropper M., Ramsay G., 2005, A\&A, 440, 185
\bibitem[\protect\citeauthoryear{Cardelli, Clayton \& Mathis}{1989}]{ccm89} Cardelli J.~A., Clayton G.~C., Mathis J.~S., 1989, ApJ, 345, 245 
\bibitem[\protect\citeauthoryear{Charbonneau}{1995}]{pikaia} Charbonneau P., 1995, ApJS, 101, 309
\bibitem[\protect\citeauthoryear{Clemens, Crain \& Anderson}{2004}]{2004SPIE.5492..331C} Clemens J.~C., Crain J.~A., Anderson R., 2004, SPIE, 5492, 331 
\bibitem[\protect\citeauthoryear{Costa \& Rodrigues}{2009}]{costa2009} Costa J.~E.~R., Rodrigues C.~V., 2009, MNRAS, 398, 240
\bibitem[\protect\citeauthoryear{Cropper}{1990}]{cropper1990} Cropper M., 1990, SSRv, 54, 195 
\bibitem[\protect\citeauthoryear{Cropper, Wu, Ramsay \& Kocabiyik}{1999}]{cropper1999} Cropper M., Wu K., Ramsay G., Kocabiyik A., 1999, MNRAS, 306, 684
\bibitem[\protect\citeauthoryear{Cruz et al.}{2018}]{cruz2018} Cruz, P., Diaz, M., Birkby, J., et al.\ 2018, MNRAS, 476, 5253 
\bibitem[\protect\citeauthoryear{Diaz \& Cieslinski}{2009}]{diaz2009} Diaz M.~P., Cieslinski D., 2009, AJ, 137, 296 
\bibitem[\protect\citeauthoryear{Diaz \& Steiner}{1994}]{diaz1994} Diaz M.~P., Steiner J.~E., 1994, A\&A, 283, 508
\bibitem[\protect\citeauthoryear{Downes et al. }{2001}]{2001PASP..113..764D} Downes R.~A., Webbink R.~F., Shara M.~M., et al.\ 2001, \pasp, 113, 764 
\bibitem[\protect\citeauthoryear{Drake et al.}{2009}]{2009ApJ...696..870D} Drake A.~J., et al., 2009, ApJ, 696, 870 
\bibitem[\protect\citeauthoryear{Drissen et al.}{1994}]{1994AJ....107.2172D} Drissen L., Shara M.~M., Dopita M., Wickramasinghe D.~T., 1994, AJ, 107, 2172 
\bibitem[\protect\citeauthoryear{Ferrario, de Martino, \& G{\"a}nsicke}{2015}]{lilia2015} Ferrario L., de Martino D., G{\"a}nsicke B.~T., 2015, SSRv, 191, 111 
\bibitem[\protect\citeauthoryear{Fossati et al.}{2007}]{2007ASPC..364..503F} Fossati L., Bagnulo S., Mason E., Landi Degl'Innocenti E., 2007, ASPC, 364, 503 
\bibitem[\protect\citeauthoryear{Gaia Collaboration et al.}{2016}]{Gaia2016} Gaia Collaboration, et al., 2016, A\&A, 595, A1
\bibitem[\protect\citeauthoryear{Gaia Collaboration et al.}{2018}]{Brown2018} Gaia Collaboration, et al., 2018, A\&A, 616, A1 
\bibitem[\protect\citeauthoryear{Hamuy et al.}{1992}]{1992PASP..104..533H} Hamuy M., Walker A.~R., Suntzeff N.~B., Gigoux P., Heathcote S.~R., Phillips M.~M., 1992, PASP, 104, 533 
\bibitem[\protect\citeauthoryear{Hayashi \& Ishida}{2014}]{hayashi2014} Hayashi T., Ishida M., 2014, MNRAS, 438, 2267
\bibitem[\protect\citeauthoryear{Henden et al.}{2015}]{2015AAS...22533616H} Henden A.~A., Levine S., Terrell D., Welch D.~L., 2015, AAS, 225, 336.16 
\bibitem[Hilditch(2001)]{Hilditch2001} Hilditch R.~W.,\ 2001, An Introduction to Close Binary Stars. Cambridge Univ. Press, Cambridge, UK, p.392
\bibitem[Iglesias-Marzoa et al. (2015)]{iglesias2015} Iglesias-Marzoa R., L{\'o}pez-Morales M., Jes{\'u}s Ar{\'e}valo Morales M., 2015, PASP, 127, 567
\bibitem[\protect\citeauthoryear{Knigge, Baraffe \& Patterson}{2011}]{Knigge11} Knigge C., Baraffe I., Patterson J., 2011, ApJS, 194, 28 
\bibitem[\protect\citeauthoryear{Livio \& Pringle}{1994}]{1994ApJ...427..956L} Livio M., Pringle J.~E., 1994, ApJ, 427, 956 
\bibitem[\protect\citeauthoryear{Magalh\~aes, Benedetti \& Roland}{1984}]{1984PASP...96..383M} Magalh\~aes A.~M., Benedetti E., Roland E.~H., 1984, PASP, 96, 383 
\bibitem[\protect\citeauthoryear{Magalh\~aes et al.}{1996}]{1996ASPC...97..118M} Magalh\~aes A.~M., Rodrigues C.~V., Margoniner V.~E., Pereyra A., Heathcote S., 1996, ASPC, 97, 118
\bibitem[\protect\citeauthoryear{McAllister, et al.}{2019}]{2019MNRAS.486.5535M} McAllister M., et al., 2019, MNRAS, 486, 5535
\bibitem[\protect\citeauthoryear{Monet}{1998}]{1998AAS...19312003M} Monet D.~G., 1998, AAS, 30, 120.03 
\bibitem[\protect\citeauthoryear{Mu{\~n}oz-Darias, Casares \& Mart{\'{\i}}nez-Pais}{2005}]{munoz-darias2005} Mu{\~n}oz-Darias T., Casares J., Mart{\'{\i}}nez-Pais I.~G., 2005, ApJ, 635, 502 
\bibitem[\protect\citeauthoryear{Oliveira et al.}{2017}]{2017AJ....153..144O} Oliveira A.~S., Rodrigues C.~V., Cieslinski D., Jablonski F.~J., Silva K.~M.~G., Almeida L.~A., Rodr{\'{\i}}guez-Ardila A., Palhares M.~S., 2017, AJ, 153, 144 
\bibitem[Pereyra (2000)]{pereyra} Pereyra A. 2000, PhD Thesis, Univ. S\~ao Paulo
\bibitem[\protect\citeauthoryear{Potter, Romero-Colmenero, Watson, Buckley \& Phillips}{2004}]{potter2004} Potter S.~B., Romero-Colmenero E., Watson C.~A., Buckley D.~A.~H., Phillips A., 2004, MNRAS, 348, 316
\bibitem[\protect\citeauthoryear{Press, Teukolsky, Vetterling \& Flannery}{1992}]{num_recipes} Press W.~H., Teukolsky S.~A., Vetterling W.~T., Flannery B.~P., 1992, Numerical recipes in FORTRAN. The art of scientific computing, 2nd ed., Cambridge Univ. Press, Cambridge
\bibitem[\protect\citeauthoryear{Rodrigues, Cieslinski \& Steiner}{1998}]{1998A&A...335..979R} Rodrigues C.~V., Cieslinski D., Steiner J.~E., 1998, A\&A, 335, 979 
\bibitem[Rosenfeld \& Kak(1982)]{rosenfeld_kak(1982)} Rosenfeld A., Kak A.~C., 1982, Computer Science and Applied Mathematics. Academic Press,  New York
\bibitem[\protect\citeauthoryear{Sarty, Saxton \& Wu}{2008}]{sarty2008} Sarty G.~E., Saxton C.~J., Wu K., 2008, Ap\&SS, 317, 239
\bibitem[\protect\citeauthoryear{Saxton et al.}{2008}]{2008A&A...480..611S} Saxton R.~D., Read A.~M., Esquej P., Freyberg M.~J., Altieri B., Bermejo D., 2008, A\&A, 480, 611 
\bibitem[\protect\citeauthoryear{Scargle}{1982}]{1982ApJ...263..835S} Scargle J.~D., 1982, ApJ, 263, 835 
\bibitem[\protect\citeauthoryear{Silva et al.}{2013}]{silva2013} Silva K.~M.~G., Rodrigues C.~V., Costa J.~E.~R., de Souza C.~A., Cieslinski D., Hickel G.~R., 2013, MNRAS, 432, 1587
\bibitem[\protect\citeauthoryear{Suleimanov, Doroshenko, Ducci, Zhukov \& Werner}{2016}]{Suleimanov_2016} Suleimanov V., Doroshenko V., Ducci L., Zhukov G.~V., Werner K., 2016, A\&A, 591, A35
\bibitem[\protect\citeauthoryear{Turnshek et al.}{1990}]{1990AJ.....99.1243T} Turnshek D.~A., Bohlin R.~C., Williamson R.~L., II, Lupie O.~L., Koornneef J., Morgan D.~H., 1990, AJ, 99, 1243 
\bibitem[\protect\citeauthoryear{Van Box Som et al.}{2018}]{vanboxsom2018} Van Box Som L., Falize {\'E}., Bonnet-Bidaud J.-M., Mouchet M., Busschaert C., Ciardi A., 2018, MNRAS, 473, 3158
\bibitem[\protect\citeauthoryear{Wade \& Horne}{1988}]{1988ApJ...324..411W} Wade R.~A., Horne K., 1988, ApJ, 324, 411 
\bibitem[\protect\citeauthoryear{Wickramasinghe \& Ferrario}{1988}]{wf1988} Wickramasinghe D.~T., Ferrario L., 1988, ApJ, 334, 412
\bibitem[\protect\citeauthoryear{Wu \& Kiss}{2008}]{2008A&A...481..433W} Wu K., Kiss L.~L., 2008, A\&A, 481, 433 
\bibitem[\protect\citeauthoryear{Wu \& Wickramasinghe}{1990}]{ww1990} Wu K., Wickramasinghe D.~T., 1990, MNRAS, 246, 686
\bibitem[\protect\citeauthoryear{Wu \& Wickramasinghe}{1992}]{ww1992} Wu K., Wickramasinghe D.~T., 1992, MNRAS, 256, 329
\bibitem[\protect\citeauthoryear{Yuasa et al.}{2010}]{yuasa2010} Yuasa T., et al., 2010, A\&A, 520, A25
\bibitem[\protect\citeauthoryear{Zorotovic, Schreiber, \& G{\"a}nsicke}{2011}]{2011A&A...536A..42Z} Zorotovic M., Schreiber M.~R., G{\"a}nsicke B.~T., 2011, A\&A, 536, A42 

\end{thebibliography}


\bsp	
\label{lastpage}
\end{document}